\documentclass[11pt,a4,twoside]{article}
\usepackage[dvips,a4paper,left=2.60cm,right=2.60cm,top=2.98cm,bottom=2.98cm,headsep=1em]{geometry}
\usepackage{fancyhdr}
\usepackage{titlesec}
\usepackage[latin1]{inputenc}
\usepackage{amsmath,amsfonts,amssymb,amstext,amsthm}
\usepackage{graphicx}
\usepackage{subcaption}
\usepackage{rotating}
\usepackage[font={small}]{caption}
\usepackage{natbib}
\usepackage{lmodern,slantsc}
\usepackage{relsize}
\usepackage[breaklinks, colorlinks=true, linkcolor=blue,
citecolor=black, urlcolor=blue, pdfborder={0 0 0},
pdfpagelabels]{hyperref}
\usepackage{tikz}
\DeclareMathAlphabet{\mathsfbf}{OT1}{cmss}{bx}{n}
\DeclareMathAlphabet{\mathmibf}{OT1}{cmr}{bx}{it}
\DeclareMathAlphabet\mathbfcal{OMS}{cmsy}{b}{n}
\def\vx{\mathbf x}
\def\vy{\mathbf y}

\def\vc{\mathbf c}

\def\vu{\mathbf u}

\def\vA{\mathbf A}

\def\vD{\mathbf D}

\def\vG{\mathbf G}

\def\v0{\boldsymbol{0}}

\def\vsigma{\boldsymbol{\sigma}}

\def\L{\langle}
\def\R{\rangle}
\newlength{\FigureHeight}
\newlength{\FigureHeightHalf}

\numberwithin{equation}{section}
\titleformat{\section}
{\large\bfseries}{\thetitle.}{0.5em}{}
\titleformat{\subsection}
{\normalfont\bfseries\filcenter}{\normalfont\bfseries\thetitle.}{0.5em}{}
\begin{document}

\title{Application of Lie-group symmetry analysis to an\\ infinite
hierarchy of differential equations at\\ the example of first
order ODEs}
\author{Michael Frewer\thanks{Email address for correspondence:
frewer.science@gmail.com}\\
\small Tr\"ubnerstr. 42, 69121 Heidelberg, Germany}
\date{{\small\today}}
\clearpage \maketitle \thispagestyle{empty}

\vspace{-2em}
\begin{abstract}
\noindent This study will explicitly demonstrate by example that
an unrestricted infinite and forward recursive hier\-archy of
differential equations must be identified as an unclosed system of
equations, despite the fact that to each unknown function in the
hierarchy there exists a corresponding determined equation to
which it can be bijectively mapped to. As a direct consequence,
its admitted set of symmetry transformations must be identified as
a weaker set of indeterminate equivalence transformations. The
reason is that no unique general solution can be constructed, not
even in principle. Instead, infinitely many disjoint and thus
independent general solution manifolds exist. This is in clear
contrast to a closed system of differential equations that only
allows for a single and thus unique general solution manifold,
which, by definition, covers all possible particular solutions
this system can admit. Herein, different first order Riccati-ODEs
serve as an example, but this analysis is not restricted to them.
All conclusions drawn in this study will translate to any first
order or higher order ODEs as well as to any PDEs.

\vspace{0.5em}\noindent{\footnotesize{\bf Keywords:} {\it Ordinary
Differential Equations, Infinite Systems, Lie Symmetries and
Equivalences, Unclosed Systems, General Solutions, Linear and Nonlinear Equations,
Initial Value Problems}}$\,$;
\end{abstract}

\section{Introduction, motivation and a brief overview what to expect\label{1}}

Infinite systems of ordinary differential equations (ODEs) appear
naturally in many applications (see e.g.
\cite{Dolph58,Bellman73,Temam97,Robinson01,Pustyl'nikov02,Haragus10}).~Mostly
they arise when regarding a certain partial differential equation
(PDE) as an ordinary differential equation on a function space.
This idea ranges back to the time of Fourier when he first
introduced his method of representing PDE solutions as an infinite
series of trigonometric basis functions, through which he
basically formed the origin of functional analysis which then a
century later was systematically developed and investigated by
Volterra, Hilbert, Riesz, and Banach, to name only a few.

The key principle behind the idea that a PDE can be viewed as a
ordinary differential equation within an infinite dimensional
space, is that every PDE by construction describes the change of a
system involving infinitely many coupled degrees of freedom, where
each degree then changes according to an ODE. In this regard, let
us sketch five simple examples to explicitly see how naturally and
in which variety such infinite dimensional ODE systems can
arise\linebreak from a PDE:

\vspace{5.5em}\noindent {\bf Example 1 (Partial
Discretization):\label{E1}} Using the numerical method of finite
differences, e.g.~on the nonlinear initial-boundary value problem
for the 2-dimensional function $u=u(t,x)$
\begin{equation}
\partial_t u=\partial^2_x u+ u^2,\;\;\text{with}\;\;
u(0,x)=\phi(x),\;\;\text{and}\;\; u(t,0)=\phi(0)=\phi(1)=u(t,1),
\label{150427:1245}
\end{equation}
in respect to the spatial variable $x$, one formally obtains the
following infinite hierarchy of coupled ODEs for the 1-dimensional
functions $u(t,k\cdot\Delta)=:u_k(t)$
\begin{equation}
\frac{du_k}{dt}=
\left(\frac{u_{k+1}-2u_k+u_{k-1}}{\Delta^2}\right)+u_k^2,\;\;
\text{for all}\;\; k=0,1,2,\dotsc, N\rightarrow\infty,
\label{150427:1243}
\end{equation}
when approximating the second partial derivative by the
equidistant central difference formula, where $\Delta=1/N$ is the
discretization size of the considered interval $0\leq x\leq 1$.
Through the given initial and boundary conditions, this system is
restricted for all $k$ and $t$ by
\begin{equation}
u_k(0)=\phi_k, \;\;\text{and}\;\; u_0(t)=\phi(0)=\phi(1)=u_N(t),
\;\;\text{for}\;\; N\rightarrow\infty.
\end{equation}
To obtain numerical results, the infinite system
\eqref{150427:1243} needs, of course, to be truncated, which then
leads to an approximation of the original PDE initial-boundary
value problem \eqref{150427:1245}. The higher $N$, or likewise,
the smaller $\Delta$, the more exact the approximation (depending
on the numerical rounding errors of the computational system,
which itself will always be limited).

\vspace{0.5em}\noindent {\bf Example 2 (Power Series
Expansion):\label{E2}} Let's assume that the linear initial value
problem
\begin{equation}
\partial_t u+\alpha_1\cdot\partial_x u +\alpha_2\cdot u=0, \;\;\text{with}\;\;
u(0,x)=\phi(x), \label{150427:1332}
\end{equation}
allows for an analytical function as solution with respect to $x$
in the interval $a\leq x\leq b$. Then the solution may be expanded
into the power series
\begin{equation}
u(t,x)=\sum_{n=0}^\infty p_n(t)\cdot x^n,
\end{equation}
which, when applied to \eqref{150427:1332}, results into the
following restricted infinite ODE system for the expansion
coefficients
\begin{equation}
\frac{d p_n}{dt}+\alpha_1\cdot (n+1)\cdot p_{n+1}+\alpha_2\cdot
p_n =0,\;\; \text{for all}\;\; n\geq 0, \;\;\text{with}\;\;
\sum_{n=0}^\infty p_n(0)\cdot x^n=\phi(x).
\end{equation}

\noindent{\bf Example 3 (Fourier's Method):\label{E3}} Consider
the initial value problem of the Burgers' equation
\begin{equation}
\partial_t u + u\cdot \partial_x u = \nu\cdot\partial_x^2 u,\;\;
\text{with}\;\; u(0,x)=\phi(x), \label{150427:1440}
\end{equation}
where $\phi(x)$ is periodic of period $2\pi$. If we are looking
for such a $2\pi$-periodic solution $u(t,x)$, we can write
\begin{equation}
u(t,x)=\sum_{n=-\infty}^\infty u_n(t)e^{i\cdot n\cdot x}.
\end{equation}
Substitution in \eqref{150427:1440} and equating the coefficients
leads to the infinitely coupled ODE relations
\begin{equation}
\frac{d u_n}{dt}+ \sum_{k+l=n} i\cdot k\cdot u_k \cdot u_l = -\nu
\cdot n^2 \cdot u_n , \;\; \text{for}\;\; (n,k,l)\in\mathbb{Z}^3,
\end{equation}
being restricted by the initial conditions $u_n(0)=\phi_n$, which
are determined by the Fourier expansion
$\phi(x)=\sum_{n=-\infty}^\infty \phi_n e^{inx}$.

\vspace{0.5em}\noindent{\bf Example 4 (Vector Space
Method):\label{E4}} This example is the generalization of the two
previous ones. Let's consider the linear initial-boundary value
problem for the diffusion equation
\begin{equation}
\partial_t u=\partial^2_x u,\;\;\text{with}\;\;
u(0,x)=\phi(x),\;\;\text{and}\;\; u(t,a)=\phi(a),\;\;
u(t,b)=\phi(b), \label{150427:1822}
\end{equation}
where $a<b$ are two arbitrary boundary points of some interval
$\mathcal{I}\subset\mathbb{R}$, which both can be also placed at
infinity. Consider further the vector space
\begin{equation}
\mathcal{V}=\Big\{\, v(x),\: \text{differentiable functions on
$\mathcal{I}$, with $v(a)=\phi(a)$, $v(b)=\phi(b)$} \,\Big\},
\label{150427:1843}
\end{equation}
with $\{w_n\}_{n=1}^\infty$ as a chosen basis, for which at this
point we do not know their explicit functional expressions, and
assume that the solution $u(t,x)$ for each value $t$ of the
initial-boundary value problem \eqref{150427:1822} is an element
of this space $\mathcal{V}$. Then we can write the solution as an
element of $\mathcal{V}$ in terms of the chosen basis vectors
\begin{equation}
u(t,x)=\sum_{n=1}^\infty u_n(t)\cdot w_n(x), \label{150427:2145}
\end{equation}
which, when inserted into \eqref{150427:1822}, induces the
following two infinite but uncoupled sets of linear eigenvalue
equations
\begin{equation}
\frac{du_n(t)}{dt}=-\lambda_n^2\cdot  u_n(t),\;\;\;\;
\frac{d^2w_n(x)}{dx^2}=-\lambda_n^2\cdot w_n(x), \;\; \text{for
all}\;\; n\geq 1, \label{150427:2042}
\end{equation}
one infinite (uncoupled) set for the expansion coefficients $u_n$,
and one infinite (uncoupled) set for the basis functions $w_n$,
where $-\lambda_n^2$ is the corresponding eigenvalue to each order
$n$. Both systems are restricted by the given initial and boundary
conditions in the form
\begin{equation}
\sum_{n=1}^\infty u_n(0)w_n(x)=\phi(x);\;\; w_n(a)=\phi(a),\;\;
w_n(b)=\phi(b),\;\;\text{for all $n\geq 1$}, \label{150429:1220}
\end{equation}
and, if the boundary values $\phi(a)$ or $\phi(b)$ are non-zero,
then the system for $u_n$ is further restricted~by
\begin{equation}
\sum_{n=1}^\infty u_n(t)=1,\;\; \text{for all possible $t$}.
 \label{150429:1221}
\end{equation}
This situation can now be generalized further when using a
different basis $\{\psi_n\}_{n=1}^\infty$ of $\mathcal{V}$
\eqref{150427:1843} than the PDE's {\it optimal} basis
$\{w_n\}_{n=1}^\infty$, which itself is defined by the PDE induced
infinite (uncoupled) ODE system \eqref{150427:2042}. For that, it
is helpful to first introduce an inner product on the outlaid
vector space $\mathcal{V}$ \eqref{150427:1843}. Given two
functions $f,g\in\mathcal{V}$, we will define the inner product
between these two functions as the following symmetric bilinear
form on $\mathcal{I}\subset\mathbb{R}$
\begin{equation}
\L f,g\R=\int_a^b f(x)\, g(x)\, dx.
\end{equation}
Now, to construct the corresponding induced infinite ODE system
for the new representation of the PDE's solution
\begin{equation}
u(t,x)=\sum_{n=1}^\infty \tilde{u}_n(t)\cdot \psi_n(x),
\label{150427:2126}
\end{equation}
it is expedient to identify the differential operator
$\partial_x^2$ in the PDE \eqref{150427:1822} as a linear operator
$A$ acting on the given vector space $\mathcal{V}$
\eqref{150427:1843}
\begin{equation}
\partial_t u=\partial_x^2 u\;\;\; \simeq\;\;\; \partial_t u=A(u),
\label{150521:2254}
\end{equation}
which then can also be written in explicit matrix-vector form as
\begin{equation}
\frac{d \tilde{\vu}}{dt}=\vA\cdot\tilde{\vu}, \label{150429:1222}
\end{equation}
where the infinite vector
$\tilde{\vu}=(\tilde{u}_1,\tilde{u}_2,\dotsc,\tilde{u}_n,\dotsc)$
is composed of the expansion coefficients of the solution vector
$u(t,x)$ \eqref{150427:2126}, and where the matrix elements of the
infinite matrix $\vA$ are given as
\begin{equation}
A_{mn}=\L \psi_m, A(\psi_n)\R=\int_a^b \psi_m(x)\, \partial_x^2
\psi_n(x)\, dx . \label{150429:1210}
\end{equation}
The explicit element values of both $\tilde{\vu}$ and $\vA$ depend
on the choice of the basis
$\{\psi_n\}_{n=1}^\infty\subset\mathcal{V}$, while the solution
vector $u(t,x)$ of the underlying PDE \eqref{150521:2254} itself
stays invariant under a change of base, i.e.,~for example, the
expansion \eqref{150427:2145} relative to the basis
$\{w_n\}_{n=1}^\infty$ represents the same solution as the
expansion \eqref{150427:2126} relative to the basis
$\{\psi_n\}_{n=1}^\infty$.

Of particular interest is now to investigate whether
$\vA=(A_{nm})_{n,m\geq 1}$ \eqref{150429:1210} represents a
symmetric matrix or not. If not, are there then any conditions
such that a symmetric matrix can be obtained? Performing a double
partial integration in \eqref{150429:1210} will yield the
expression
\begin{align}
A_{mn} &= \int_a^b \psi_m(x)\,
\partial_x^2 \psi_n(x)\, dx\nonumber\\[0.25em]
& = \int_a^b \psi_n(x)\, \partial_x^2 \psi_m(x)\, dx
+\Big[\psi_m(x)\partial_x
\psi_n(x)-\psi_n(x)\partial_x\psi_m(x)\Big]_{x=a}^{x=b}\nonumber\\[0.25em]
& = A_{nm}+\Big[\psi_m(x)\partial_x
\psi_n(x)-\psi_n(x)\partial_x\psi_m(x)\Big]_{x=a}^{x=b},
\label{150429:1132}
\end{align}

\noindent which implies that in order to obtain a symmetric
matrix, we have to choose the boundary values $\phi(a)$ and
$\phi(b)$ such that they satisfy the symmetry relation
\begin{equation}
\Big[\psi_m(x)\partial_x
\psi_n(x)-\psi_n(x)\partial_x\psi_m(x)\Big]_{x=a}^{x=b}=0,\;\;\text{for
all $\, n,m\geq 1$}. \label{150429:1157}
\end{equation}
For example, the simplest choice is to enforce vanishing boundary
values $\phi(a)=\phi(b)=0$. This will satisfy the condition
\eqref{150429:1157} {\it independently} on the choice of the basis
functions, with the advantageous effect then that the matrix $\vA$
is a real symmetric matrix to {\it any} chosen basis of the vector
space $\mathcal{V}$ \eqref{150427:1843}. In particular, since the
eigenvectors of a real symmetric matrix are orthogonal, the matrix
elements $A_{mn}$ for the optimal basis $\{w_n\}_{n=1}^\infty$,
according to \eqref{150427:2042}, form the diagonal matrix
\begin{equation}
A_{mn}=-\lambda^2_n \cdot \|w_n\|^2\,\delta_{mn}=\int_a^b w_m(x)\,
\partial_x^2 w_n(x)\, dx,
\end{equation}
where no summation over the repeated index $n$ is implied. Hence,
if we choose the alternative basis $\{\psi_n\}_{n=1}^\infty$ such
that it's orthonormal\footnote[2]{If $\{\psi_n\}_{n=1}^\infty$ is
an orthonormal basis of $\mathcal{V}$, then each $\psi_n$ must be
collinear to $w_n$, in particular $\psi_n=w_n/\|w_n\|$.}
\begin{equation}
\L \psi_m, \psi_n\R = \delta_{mn},
\end{equation}
then, according to \eqref{150429:1220} and \eqref{150429:1221},
the PDE's induced infinite ODE system \eqref{150429:1222} will
only be restricted by the initial conditions
\begin{equation}
\tilde{u}_n(0)=\int_a^b \phi(x)\, \psi_n(x)\, dx,\;\; \text{for
all} \;\; n\geq 1.
\end{equation}

\vspace{11.5em}\noindent{\bf Example 5 (Method of
moments):\label{E5}} Consider the initial value problem (Cauchy
problem) of the more generalized linear diffusion equation
\begin{equation}
\partial_t u = a\cdot\partial^2_x u + b\cdot x\cdot\partial_x u +
(b+c\cdot x^2)\cdot u, \;\;\text{with}\;\; u(0,x)=\phi(x).
\label{150430:1313}
\end{equation}
If one is interested in the moments
\begin{equation}
u_n(t)=\int_{-\infty}^\infty x^n\cdot u(t,x)\, dx,\;\; n\geq 0,
\end{equation}
by multiplying the PDE \eqref{150430:1313} with $x^n$ and
integrating over~$\mathbb{R}$, and if one assumes that partial
integration is justified, i.e. when assuming the natural boundary
conditions
\begin{equation}
\lim_{x\rightarrow \pm\infty} u(t,x)=0,\quad \lim_{x\rightarrow
\pm\infty} \partial_x u(t,x)=0,
\end{equation}
then one obtains the following infinite system of coupled ODEs for
all $n\geq 0$
\begin{equation}
\frac{d u_n}{dt} = a\cdot n\cdot(n-1)\cdot u_{n-2}-b\cdot n\cdot
u_n +c\cdot u_{n+2},\;\; \text{with}\;\;
u_n(0)=\int_{-\infty}^\infty x^n\cdot \phi(x)\, dx.\;\;
\label{150501:1415}
\end{equation}
Note that for $c=0$ the solution $u=u(t,x)$ of \eqref{150430:1313}
has the property of a probability measure, for example in that it
can be interpreted as a probability density of a particle
undergoing Brownian motion.~Only for $c=0$ the PDE
\eqref{150430:1313} attains the structure of a Fokker-Planck
equation with the drift coefficient $D_1(t,x)=-b\cdot x$ and the
diffusion coefficient $D_2(t,x)=a>0$
\begin{equation}
\partial_t u = \partial_x \big(-D_1\cdot u+D_2\cdot\partial_x
u\big), \label{150504:1046}
\end{equation}
which describes the time evolution of the probability distribution
$u(t,x)\geq 0$ such that no probability is lost, i.e. conserved
for all $t\geq 0$
\begin{equation}
\int_{-\infty}^\infty u(t,x)\, dx =1, \label{150501:1337}
\end{equation}
due to the defining structure of equation \eqref{150504:1046}
being a conservation law with the probability current $J=-D_1\cdot
u+D_2\cdot\partial_x u$. Hence, for $c=0$ the corresponding
infinite ODE system \eqref{150501:1415} is thus further
(automatically) restricted by $u_0(t)=1$ for all times $t\geq 0$.

However, if $c\neq 0$ then the zeroth moment \eqref{150501:1337}
in general is not constant in time; it will rather evolve
according to some prescribed 'normalization' function $N(t)$
\begin{equation}
\int_{-\infty}^\infty u(t,x)\, dx =N(t). \label{150504:1112}
\end{equation}
To exogenously enforce a certain function $N(t)=N_0(t)$ as an
additional (non-local) boundary condition onto the Cauchy problem
\eqref{150430:1313} would result into an overdetermined system,
for which no solutions may exist. The Cauchy problem itself, e.g.
of equation \eqref{150430:1313}, is well-posed and allows, up to a
normalization constant, for a unique solution. The normalization
constant can be fixed by posing at $t=0$ a normalized initial
condition, e.g. $\int_{-\infty}^\infty
\phi(x)dx=N(0)=1$.\linebreak Note that although such integral
overdetermination conditions as \eqref{150504:1112} are widely
used to study associated inverse problems, e.g. to find
corresponding heat sources and diffusion coefficients (see e.g.
\cite{Dehghan07,Kanca12,Hazanee13}), we will not consider them
here.

\pagebreak[4] These five examples discussed above show that there
is a multitude of possibilities in how a PDE can induce an ODE in
an infinite dimensional space.~If the PDE is restricted by initial
or boundary conditions they are transcribed to the infinite system
accordingly. Mostly, only the initial conditions get directly
transferred, while the boundary conditions are only needed as
auxiliary conditions to actually perform the reduction process,
e.g. as in the case of Example~\hyperref[E4]{4}. Note that all
examples only considered the reduction of $(1+1)$-dimensional
parabolic PDEs, but it's obvious that this concept extends to any
type of PDEs of any dimension. The result is then not a single
infinite system, but rather a collective hierarchy of several
infinite systems of coupled~ODEs.

\vspace{-0.177em} As all examples showed, it should be clear that
the associated infinite system of ODEs is {\it not} identical to
the PDE. It only represents a reduction of the PDE, since always a
certain Ansatz of the PDE's solution manifold has to be made in
order to obtain its associated infinite ODE system. Stated
differently, the PDE operates on a higher level of abstraction
than its induced infinite system of (lower level) ODEs, which,
although infinite dimensional, nevertheless depends on assumptions
and in particular on the choice of the reduction method used. That
is, a {\it single} PDE can always be reduced to a multitude of
functionally and structurally {\it different} infinite systems of
ODEs depending on the choice of method. This insight can be
transferred to differential equations which operate on an even
higher abstraction level than PDEs, e.g.~so-called functional
equations which involve functional derivatives. Then an infinite
hierarchy of PDEs instead of ODEs takes the place of the reduced
system.

A natural question which arises is whether the infinite set of
reduced equations is easier to analyze than the original PDE? In
general, the answer to this question is "no". However, if the
infinite system is truncated and approximated to a low-dimensional
form, then often qualitative analysis is possible, and useful
insights into the dynamics of the original system can be obtained.
Also from a numerical point of view many interesting stability
questions arise when the system is truncated, because, in order to
obtain numerical results from an infinite system, some method of
truncation must be employed. Surely, the quality of the subsequent
approximation towards a consistent finite dimensional system
strongly depends on this method in how the system was truncated,
which is part of the theories of closure and
differential~approximation.

The formal mathematical environment to study and analyze an
infinite (non-truncated) sequence of differential equations is set
by the infinite-dimensional theory of Banach spaces. Questions
regarding existence and uniqueness of solutions can only be
properly dealt with from the perspective of a Banach space in
defining and constructing appropriate functional norms. Such
systematic investigations, however, are beyond the scope of this
article; for that, the rich literature on this topic has to be
consulted (see e.g.
\cite{Tikhonov34,Valeev74,Deimling77,Samoilenko03,Hajek10,Fabian11}).
Instead, we will only make a small excursion into the uniqueness
issue of these solutions when restricting the infinite system by a
sufficient set of initial conditions, but only to show where still
the problems lie and not on how to solve these problems.

The main focus of this article will be based on the {\it
unrestricted} infinite set of ODEs, and to primarily study the
{\it general} solutions they admit. By taking the perspective of a
Lie group based symmetry analysis
\citep{Stephani89,Fushchich93,Olver93,Ibragimov94,Bluman10}, we
can demonstrate by example that eventually any unrestricted
infinite set of differential equations, which is based on a {\it
forward} recurrence relation, must be identified as an unclosed
and thus indeterminate system, although, in a formal one-to-one
manner, one can associate to each equation in the hierarchy a
corresponding unknown function. As a consequence, such unclosed
differential systems do not allow for the construction of a unique
{\it general} solution.~Any desirable {\it general} solution can
be generated. The side-effect of this result is that each symmetry
transformation then only acts in the weaker sense of an
equivalence transformation
\citep{Ovsiannikov82,Meleshko96,Ibragimov04,Vaneeva14}.~Such an
identification is necessary in order to allow for a consistent
invariance analysis among an infinite set of differential
equations.

\vspace{-0.177em} At first sight it may seem to be a trivial
observation that an {\it unrestricted} infinite set of ODEs has
the property of an underdetermined system.~Because if it
represents a specific reduction\linebreak of an {\it unrestricted}
PDE, i.e.~of a PDE which is not accompanied by any initial or
boundary\linebreak conditions, its general solution is only unique
up to certain integration functions.~And since this arbitrariness
on the higher abstraction level of the PDE is transferred down to
the lower abstraction level of the reduced ODE system, it is not
surprising that the latter system is somehow arbitrary as well.
But, by closer inspection there is no one-to-one correspondence,
because, for example, for any evolutionary PDE with fixed spatial
boundary conditions, the degree of arbitrariness in its general
solution only depends on the order of the time derivative, which
in turn is directly linked to the number of initial condition
functions needed to generate a unique solution from the general
one.~However, for its reduced ODE system the degree of
arbitrariness is differently larger in that it not only depends on
the temporal differential order as the underlying PDE does, but
also, additionally, on the direction and the order of the spatial
recurrence relation which this system inherently defines. For
example, if the recurrence relation is a {\it forward} recurrence
of order one, then, independent of the temporal differential
order, one unknown function anywhere in the ODE hierarchy can be
specified freely; if its a {\it forward} recurrence of order two,
then two unknown functions can be specified freely, and so on.
This freedom in choice has no correspondence on the higher
abstraction level of the PDE. Appendix \ref{A1} and \ref{A2}
provide a preview demonstration of these statements by considering
again Example \hyperref[E5]{5}.\linebreak In Section \ref{3.1} and
Section \ref{3.2} this insight will then be investigated in more
detail by also involving different examples.

In particular, it was exactly Example \hyperref[E5]{5} with its
properties discussed in Appendix \ref{A1} and \ref{A2} which
motivated this study. The inherent principle that the (higher
abstraction level) PDE \eqref{150430:1313} represents a closed
system while the correspondingly reduced (lower abstraction level)
ODE system of its moments \eqref{150501:1415}, although being
infinite in dimension, constitutes an unclosed system {\it if} the
recurrence is of forward direction (see Appendix \ref{A2}),
obviously transfers to an even higher abstraction level of
description, as seen, for example, when formulating the
statistical description of Navier-Stokes turbulence. There the
functional Hopf equation formally serves as the (higher
abstraction level) {\it closed} equation while its correspondingly
induced (lower abstraction level) infinite Friedmann-Keller PDE
system of multi-point moments is unclosed; for more details on
this issue, see \cite{Frewer14.1} and \cite{Frewer15.1}.

Part of the current study is to mathematically clarify this point
in statistical turbulence research, namely where any formally {\it
closed} set of equations which operates on a higher statistical
level always induces an {\it unclosed} infinite system on the
lower statistical level of the {\it moment} equations, and, where
thus, both levels of description are not equivalent. The reason
for this is that due to the nonlinearity of the Navier-Stokes
problem a {\it forward} recurrence relation is always generated on
the lower abstraction level of the statistical moments (in the
sense similar to the problem demonstrated in Appendix \ref{A2}),
turning thus the corresponding infinite Friedmann-Keller PDE
hierarchy inherently into an unclosed system. It is necessary to
clarify this point, because it seems that in the relevant
literature on turbulence there still exists a misconception on
this issue, in particular in the studies of
Oberlack~et~al.~\citep{Oberlack10,Oberlack13.1,Oberlack14,Oberlack14.1,Oberlack14.2}.
A detailed discussion on this misconception is given in
\cite{Frewer15.1}.

The paper is organized as follows: Section \ref{2} first considers
a single (closed) ODE to define the concept of a {\it unique}
general solution manifold from the perspective of an invariance
analysis. In Section \ref{3} this concept will be applied to an
(unclosed) infinite ODE system based on a {\it forward} recurrence
relation. Both a linear (Section \ref{3.1}) as well as a nonlinear
system (Section \ref{3.2}) will be investigated, which both stem
as special cases from a generalized hierarchy of first order
Riccati-ODEs. To which higher level PDE this infinite ODE system
belongs to is an inverse problem, which will not be investigated
since it's clearly beyond the scope of this article. Based on the
results presented herein, we can conclude that any unrestricted
infinite system which follows a forward recurrence relation must
be identified as an unclosed system, which then, as consequence,
only leads to non-unique general solution manifolds and which,
instead of symmetry transformations, only admits the weaker
equivalence transformations. This twofold conclusion is
independent of whether an infinite hierarchy of ODEs or whether an
infinite hierarchy of PDEs is considered.

\section{Lie-point symmetries and general solution of a single Riccati-ODE\label{2}}

In general a Riccati equation is any first-order ODE that exhibits
a quadratic nonlinearity of the form
\begin{equation}
y^\prime(x)=q_0(x)+q_1(x)y(x)+q_2(x)y^2(x), \label{150401:1222}
\end{equation}
with $q_i(x)$ being arbitrary functions (see e.g.
\cite{Reid72}).\footnote[2]{Note that the nonlinear Riccati-ODE
\eqref{150401:1222} can always be reduced to a linear ODE of
second order by making use of the transformation $y(x)=-z^\prime
(x)/(q_2(x)\cdot z(x))$.} In this section, however, we only want
to consider the following specific Riccati-ODE
\begin{equation}
y^\prime -\frac{y}{x}-\frac{y^2}{x^3}=0, \label{150323:1732}
\end{equation}
which is also categorized as a specific Bernoulli differential
equation of the quadratic rank (see e.g. \cite{Parker13}). Its
unique general solution\footnote[3]{Transforming the nonlinear ODE
\eqref{150323:1732} according to $y(x)=x/z(x)$ will reduce it to a
linear ODE of first order which can be solved then by a simple
integration.} is given by
\begin{equation}
y(x)=\frac{x^2}{1+c\cdot x},\quad c\in\mathbb{R},
\label{150323:2207}
\end{equation}
involving a single free integration parameter $c$. Since
\eqref{150323:1732} is a {\it single first order} ODE it has the
special property of admitting an infinite set of Lie-point
symmetries (see e.g.~\cite{Stephani89,Bluman96,Ibragimov04}). The
symmetries are generated by the tangent field
$X=\xi(x,y)\partial_x +\eta(x,y)\partial_y$, which, in the
considered case \eqref{150323:1732}, satisfies the following
underdetermined relation for the infinitesimals $\xi=\xi(x,y)$ and
$\eta=\eta(x,y)$:
\begin{align}
\!\!\!\!\! 0= &\; \xi\cdot(3y^2x^2+yx^4)-\eta\cdot (2yx^3+x^5)\nonumber\\
& -\partial_x\xi\cdot (y^2x^3+yx^5)-\partial_y\xi\cdot
(y^4+2y^3x^2+y^2x^4)+\partial_x\eta\cdot (x^6)
+\partial_y\eta\cdot (y^2x^3+yx^5). \label{150323:1836}
\end{align}
Note that in constructing the general solution of equation
\eqref{150323:1836} only one function can be chosen arbitrarily,
either $\xi$ or $\eta$, but not both. Without restricting the
general case, we will choose $\xi$ as the free infinitesimal,
which, once chosen in \eqref{150323:1836}, then uniquely fixes the
second infinitesimal $\eta$. For the present, it is sufficient to
only consider monomials in the normalized form $x^n$ with $n\geq
0$ as a functional choice for $\xi$. According to
\eqref{150323:1836}, the corresponding tangent field $X$ up to
order $n$ is then given by
\begin{align} \mathsf{T}_n: & \;\;\;
X_n =x^n\partial_x +
\left[x^{n+2}\cdot\left(\frac{y}{x^3}+\frac{y^2}{x^5}\right)
+F_n\!\!\left(\frac{y-x^2}{yx}\right)\cdot\frac{y^2}{x}\right]\partial_y,
\quad n\geq 0, \label{150323:1920}
\end{align}
where the $F_n$ are arbitrary integration functions with argument
$(y-x^2)/yx$. For the sake of simplicity it is convenient to
choose these functions such that for each order $n$ the lowest
degree of complexity is achieved. For example, for the first four
elements in this chain \eqref{150323:1920} we choose the $F_n$
such that
\begin{equation}
\left. \begin{aligned} \mathsf{T}_0: & \;\;\;
X_0=\partial_x+\left(\frac{3y}{x}-x\right)\partial_{y},\\[0.5em]
\mathsf{T}_1: & \;\;\; X_1=x\partial_x+2y\partial_{y},\\[0.75em]
\mathsf{T}_2: & \;\;\; X_2=x^2\partial_x+ xy\partial_y,\\[0.75em]
\mathsf{T}_3: & \;\;\;
X_3=x^3\partial_x+\left(y^2+yx^2\right)\partial_y,
\end{aligned}
~~~ \right \} \label{150323:1919}
\end{equation}

\noindent which, according to Lie's central theorem (see e.g.
\cite{Bluman96}), are equivalent to the 1-parameter symmetry group
transformations
\begin{equation}
\left. \begin{aligned} \mathsf{T}_0: & \;\;\;
\tilde{x}=x+\varepsilon_0,\;\;\;
\tilde{y}=\left(\frac{y-x^2}{x^3}+\frac{1}{x+\varepsilon_0}\right)
(x+\varepsilon_0)^3,\\[0.5em]
\mathsf{T}_1: & \;\;\; \tilde{x}=e^{\varepsilon_1} x,\;\;\;
\tilde{y}=e^{2\varepsilon_1}y,\\[0.75em]
\mathsf{T}_2: & \;\;\; \tilde{x}=
\frac{x}{1-\varepsilon_2\, x},\;\;\; \tilde{y}=\frac{y}{1-\varepsilon_2\, x},\\[0.75em]
\mathsf{T}_3: & \;\;\; \tilde{x}=\frac{x}{\sqrt{1-2\varepsilon_3\,
x^2}},\;\;\; \tilde{y}=\frac{yx^2}{y\cdot (1-2\varepsilon_3\,
x^2)-(y-x^2)\cdot\sqrt{1-2\varepsilon_3\, x^2}}.
\end{aligned}
~~~ \right \} \label{150323:2000}
\end{equation}

\noindent By construction each of the above transformations leaves
the considered differential equation
\eqref{150323:1732}~invariant, i.e.~when transforming
\eqref{150323:1732} according to one of the transformations
\eqref{150323:2000} will thus result into the invariant form
\begin{equation}
\tilde{y}^\prime
-\frac{\tilde{y}}{\tilde{x}}-\frac{\tilde{y}^2}{\tilde{x}^3}=0.
\label{150323:2126}
\end{equation}
Note that for any point transformations, as in the case
\eqref{150323:2000}, the transformation for the first order
ordinary derivative is induced by the relation
\begin{equation}
\tilde{y}^\prime=\frac{d
\tilde{y}}{d\tilde{x}}=\frac{\frac{\partial \tilde{y}}{\partial
x}dx+\frac{\partial \tilde{y}}{\partial y}dy}{\frac{\partial
\tilde{x}}{\partial x}dx+\frac{\partial \tilde{x}}{\partial y}dy}
=\frac{\frac{\partial \tilde{y}}{\partial x}+\frac{\partial
\tilde{y}}{\partial y}y^\prime}{\frac{\partial \tilde{x}}{\partial
x}+\frac{\partial \tilde{x}}{\partial
y}y^\prime}=\left(\frac{\partial\tilde{x}}{\partial x}\right)^{-1}
\left(\frac{\partial \tilde{y}}{\partial x}+\frac{\partial
\tilde{y}}{\partial y}y^\prime\right), \label{150404:1245}
\end{equation}
where the last equality only stems from the fact that
$\partial{\tilde{x}}/\partial y =0$ for all transformations
\eqref{150323:2000}.

Now, if the general solution \eqref{150323:2207} would not be
known beforehand, then the symmetries \eqref{150323:2000} can be
used to construct it. For any first order ODE, as in the present
case for \eqref{150323:1732}, at least one symmetry is necessary
to determine its general solution, which, for example, can be
achieved by making use of the method of canonical variables (see
e.g. \cite{Stephani89}). But, instead of performing this
construction, the opposite procedure will be investigated, namely
to validate in how far the function \eqref{150323:2207} represents
a unique general solution\footnote[2]{By definition a unique
general solution of a differential equation or a system of
differential equations should cover {\it all} particular (special)
solutions this system can admit. In other words, every special
solution that can be constructed must be covered through this
general solution by specifying a corresponding initial condition,
otherwise the given general solution is not complete or unique.}
of \eqref{150323:1732} when transforming it according to the
symmetries \eqref{150323:2000}. The result to expect is that if
function \eqref{150323:2207} represents the unique general
solution of \eqref{150323:1732}, then it either must map to
\begin{equation}
\tilde{y}(\tilde{x})=\frac{\tilde{x}^2}{1+\tilde{c}\cdot
\tilde{x}},\quad \tilde{c}\in\mathbb{R}, \label{150323:2318}
\end{equation}
with a new free transformed parameter $\tilde{c}=\tilde{c}(c)$ as
a function of the old untransformed parameter $c$, or it must
invariantly map to
\begin{equation}
\tilde{y}(\tilde{x})=\frac{\tilde{x}^2}{1+c\cdot \tilde{x}},\quad
c \in\mathbb{R}, \label{150323:2319}
\end{equation}
with an unchanged free parameter $c$ before and after the
transformation. The reason is that since all transformations
\eqref{150323:2000} form true symmetries, which map solutions to
new solutions of the underlying differential equation
\eqref{150323:1732}, any solution which forms a unique general
solution of this equation can thus only be mapped into itself,
either into the non-invariant form \eqref{150323:2318} or into the
invariant form \eqref{150323:2319}, because no other functionally
independent solution exists to which the symmetries can map to. If
this is not the case, we then have to conclude that either the
considered transformation is not a symmetry transformation or that
the given solution is not the general solution.

Hence, since we definitely know that \eqref{150323:2207} is the
unique general solution of equation \eqref{150323:1732}, which
again admits the symmetries \eqref{150323:2000}, these symmetry
transformations only need to be classified into two categories,
namely into those which reparametrize the general solution
\eqref{150323:2318} and into those which leave it invariant
\eqref{150323:2319}. For example, the symmetries $\mathsf{T}_0$,
$\mathsf{T}_1$ and $\mathsf{T}_2$ reparametrize the general
solution \eqref{150323:2207} with
$\tilde{c}=c/(1-c\,\varepsilon_0)$,
$\tilde{c}=e^{-\varepsilon_1}c$, and $\tilde{c}=c+\varepsilon_3$
respectively, while symmetry $\mathsf{T}_3$ keeps it invariant.
This game can then be continued for all higher orders of $n$ in
\eqref{150323:1920}, or even for any other functionally different
symmetry using the general determining relation
\eqref{150323:1836}.

To conclude this section, it is helpful to formalize the above
insights: Let $f_\lambda$ formally be a parameter dependent
solution of a differential equation $E$, and $\mathsf{S}$ any
transformation which leaves this differential equation invariant
$\mathsf{S}(E)=E$. The transformation $\mathsf{S}$ on $f_\lambda$
is called a reparametrization if
$\mathsf{S}(f_\lambda)=f_{s(\lambda)}$, which includes the special
case of an invariant transformation if $s(\lambda)=\lambda$, where
the parameter mapping $s$ is induced by the variable
mapping~$\mathsf{S}$.\linebreak Then, based on these conditions,
the following two statements are equivalent:
\begin{equation}
\left .
\begin{aligned}
\quad\quad\;\text{$f_\lambda$ is a unique general solution}\quad &
\Rightarrow & \;\,
\mathsf{S}(f_\lambda)=f_{s(\lambda)},\hspace{3.68cm}\\[0.5em]
\mathsf{S}(f_\lambda)\neq f_{s(\lambda)}\quad & \Rightarrow & \;\,
\text{$f_\lambda$ is not a unique general solution},
\end{aligned}
~~~ \right\}\label{150425:1121}
\end{equation}

\noindent where in each case the opposite conclusion is, of
course, {\it not} valid, i.e.
\begin{equation}
\left .
\begin{aligned}
\mathsf{S}(f_\lambda)=f_{s(\lambda)}\quad & \nRightarrow & \;\,
\text{$f_\lambda$ is a unique general solution},\\[0.5em]
\text{$f_\lambda$ is not a unique general solution}\quad &
\nRightarrow & \;\,\mathsf{S}(f_\lambda)\neq
f_{s(\lambda)}.\hspace{2.95cm}
\end{aligned}
~~~~~~~~ \right\}\!\!\!\label{150425:1142}
\end{equation}

\section{Lie-point symmetries and general solution for an infinite
system of ODEs\label{3}}

Let's consider the following {\it unrestricted} infinite and {\it
forward} recursive hierarchy of ordinary differential equations
based on the Riccati ODE \eqref{150401:1222}
\begin{equation}
y_n^\prime(x)-q_0(x)-q_1(x)y_n(x)=q_2(x)y_{n+1}^2(x)+q_3(x)y_{n+1}(x),\quad
n=1,2,3,\,\dotsc \label{150401:1710}
\end{equation}
A solution of such a system is defined as an infinite set of
functions $\{y_1(x),y_2(x),\ldots,y_n(x),\ldots\}$ for which all
the equations of the system hold identically. Without restricting
the general case, we will consider two specifications: a linear
and a nonlinear one.

\subsection{Infinite linear hierarchy of first order ODEs\label{3.1}}

In this section we will consider the linear specification
$q_0=q_1=q_2=0,q_3=-1$ of \eqref{150401:1710}
\begin{equation}
y_n^\prime = -y_{n+1},\quad n=1,2,3,\,\dotsc ,\label{150401:1859}
\end{equation}
which also can be equivalently written in vector form as
\begin{equation}
\vy^\prime=-\vA\cdot\vy,\label{150401:1830}
\end{equation}
where $\vA$ is the infinite but bounded bi-diagonal matrix
\begin{equation*}
\vA=\begin{pmatrix}
\,\, 0 & 1 & 0 & 0 & 0 & \cdots & 0 & \cdots\,\, \\
\,\, 0 & 0 & 1 & 0 & 0 & \cdots & 0 & \cdots\,\, \\
\,\, 0 & 0 & 0 & 1 & 0 & \cdots & 0 & \cdots\,\, \\
\,\, \vdots & \vdots & \vdots & \ddots & \ddots & \vdots & \vdots
& \vdots\,\,
\end{pmatrix},
\end{equation*}
along with the infinite dimensional solution vector
$\vy^T=(y_1,y_2,y_3,\dots ,y_n,\dots)$ of \eqref{150401:1830}.
Naively one would expect that the unique general solution to
\eqref{150401:1830} is given by\footnote[2]{The operator
$e^{-x\vA}$ is called the flow of the differential equation
\eqref{150401:1830}, as it takes the initial state $\vy=\vc$ at
$x=0$ into the new state $\vy=e^{-x\vA}\cdot \vc$ at position
$x\neq 0$. If \eqref{150401:1830} represents an evolution equation
with its forward marching time $t\geq 0$ as the independent
variable, then the set of all operator elements $e^{-t\vA}$ only
forms a semi-group. The operator $\vA$ is then said to be the
infinitesimal generator of this semi-group.}
\begin{equation}
\vy =e^{-x\vA}\cdot \vc, \label{150401:1848}
\end{equation}
where $\vc$ is the infinite dimensional integration constant
$\vc^T=(c_1,c_2,c_3,\dots c_n,\dots)$. When evolving the
exponential function into its power series with its infinite
radius of convergence, the general solution \eqref{150401:1848}
can be equivalently written as
\begin{equation}
\left.
\begin{aligned}
y_1(x)=c_1-c_2\cdot x+ \frac{1}{2!}\, c_3\cdot x^2 -
\frac{1}{3!}\, c_4\cdot x^3 + \cdots \\
y_2(x)=c_2-c_3\cdot x+ \frac{1}{2!}\, c_4\cdot x^2 -
\frac{1}{3!}\, c_5\cdot x^3 + \cdots \\
y_3(x)=c_3-c_4\cdot x+ \frac{1}{2!}\, c_5\cdot x^2 -
\frac{1}{3!}\, c_6\cdot x^3 + \cdots \\
\vdots\hspace{3.2cm}\vdots\hspace{3cm}\vdots\hspace{0.15cm}
\end{aligned}
~~~ \right \} \label{150401:1909}
\end{equation}
or compactly as
\begin{equation}
y_n(x)=\sum_{k=0}^\infty c_{n+k}\frac{(-1)^k}{k!}\, x^k,\quad
n=1,2,3,\,\dotsc ,\label{150401:1925}
\end{equation}
which, naively considered, might then serve as the unique general
solution for \eqref{150401:1859}. Of course, the precondition for
it is that for any given initial condition the constant component
values of the infinite dimensional vector $\vc$ must be given such
that the matrix product \eqref{150401:1848} is converging. That
\eqref{150401:1848}, or equivalently \eqref{150401:1925},
represents a {\it general} solution to \eqref{150401:1830} is
obvious, because to every (first order) differential equation of
the hierarchy \eqref{150401:1859} one can associate a solution
$y_n$ involving a free integration parameter $c_n$.

Now, let us see in how far the general solution
\eqref{150401:1848} represents a {\it unique} general solution of
\eqref{150401:1830}. For that we first consider one of the
equations' scaling symmetries admitted by \eqref{150401:1830}
\begin{equation}
\mathsf{L}_1: \;\;\; \tilde{x}=e^{-\varepsilon}x,\;\;\;
\tilde{\vy}=\vD(\varepsilon)\cdot\vy,\label{150403:1113}
\end{equation}
where $\vD$ is the infinite diagonal matrix
\begin{equation*}
\vD(\varepsilon)=\begin{pmatrix}
\,\, e^{\varepsilon} & 0 & 0 & 0 & \cdots\,\, \\
\,\, 0 & e^{2\varepsilon} & 0 & 0 & \cdots\,\, \\
\,\, 0 & 0 & e^{3\varepsilon} & 0  & \cdots\,\, \\
\,\, \vdots & \vdots & \cdots & \ddots & \vdots\,\,
\end{pmatrix}.
\end{equation*}
Transforming the general solution \eqref{150401:1848} according to
$\mathsf{L}_1$ \eqref{150403:1113} will map the solution up to a
reparametrization in the integration constant $\vc\mapsto
\tilde{\vc}$ into itself$\,$\footnote[2]{To obtain the second
relation in \eqref{150403:1149} one has to use the non-commutative
property $\vD\cdot \vA^n=e^{-n\varepsilon} (\vA^n\cdot \vD)$.}
\begin{equation}
\tilde{\vy}=\vD(\varepsilon)\cdot
\Big(e^{-e^{\varepsilon}\tilde{x}\vA}\cdot\vc\Big)
=e^{-\tilde{x}\vA} \cdot \big(\vD(\varepsilon)\cdot\vc\big)=
e^{-\tilde{x}\vA}\cdot\tilde{\vc}.\label{150403:1149}
\end{equation}
That means, regarding symmetry transformation $\mathsf{L}_1$
\eqref{150403:1113} the general solution \eqref{150401:1848}
represents itself as unique general solution indeed. But this is
no longer the case if we consider for example the following
symmetry transformation
\begin{align}
\mathsf{L}_2: \;\;\; &
X=\xi(x,y_1,y_2,\dotsc)\,\partial_x+\sum_{n=1}^\infty\eta_n(x,y_1,y_2,\dotsc)\,\partial
_{y_n},\label{150403:1420}\\
& \;\;\text{with}\;\;\xi=x^2,\;\;\text{and}\;\;
\eta_n=y_n+(n-1)(n-2)\, y_{n-1}-2\,(n-1)\, x\, y_{n},
\hspace{3.35cm}\nonumber
\end{align}
which in global form reads as (see derivation
\eqref{150404:1253}-\eqref{150518:1452})
\begin{equation}
\left.
\begin{aligned}
\negthickspace\negthickspace\negthickspace\!\mathsf{L}_2: \;\;\; &
\tilde{x}=\frac{x}{1-\varepsilon
x},\;\;\;\tilde{y}_1=e^{\varepsilon}y_1,\\
& \tilde{y}_n=\, \sum_{k=1}^{n-1}
B_{n,k}\,\varepsilon^{n-k-1}(1-\varepsilon x)^{n+k-1}\,
e^{\varepsilon}\, y_{k+1},\;\;\text{for all}\;\; n\geq
2,\qquad\qquad\qquad\quad\;\,
\end{aligned}
~~~ \right \} \label{150403:1804}
\end{equation}
where e.g.~the first three explicit elements in this hierarchy are
given as
\begin{equation}
\left.
\begin{aligned}
\negthickspace\negthickspace\negthickspace\,\!\! \mathsf{L}_2:
\;\;\; & \tilde{x}=\frac{x}{1-\varepsilon
x},\;\;\;\tilde{y}_1=e^{\varepsilon}y_1,\\ &
\tilde{y}_2=(1-\varepsilon x)^2\, e^\varepsilon y_2,\;\;\;
\tilde{y}_3=2\varepsilon(1-\varepsilon x)^3\, e^\varepsilon y_2 +
(1-\varepsilon x)^4\, e^\varepsilon y_3,\;\;\; \cdots
\qquad\quad\;\;\;\;\;\,
\end{aligned}
~~~ \right \} \label{150403:1855}
\end{equation}
Because when transforming the general solution \eqref{150401:1848}
according to the above symmetry transformation
\eqref{150403:1804}, which in matrix-vector form reads as
\begin{equation}
\mathsf{L}_2: \;\;\; \tilde{x}=\frac{x}{1-\varepsilon
x},\;\;\;\tilde{\vy}=\vG(x,\varepsilon)\cdot\vy,
\label{150407:1002}
\end{equation}
where $\vG$ is the infinite group matrix
\begin{equation*}
\vG(x,\varepsilon)=e^{\varepsilon}\begin{pmatrix}
\,\, 1 & 0 & 0 & 0 & 0 & \cdots\,\, \\
\,\, 0 & B_{2,1}\varepsilon^0(1-\varepsilon x)^2 & 0 & 0 & 0 & \cdots\,\, \\
\,\, 0 & B_{3,1}\varepsilon^1(1-\varepsilon x)^3
& B_{3,2}\varepsilon^{0}(1-\varepsilon x)^4  & 0 & 0 &\cdots\,\, \\
\,\, \vdots & \vdots & \vdots & \ddots & \vdots & \vdots \,\, \\
\;\; 0 & B_{n,1}\varepsilon^{n-2}(1-\varepsilon x)^n & \cdots &
B_{n,n-1}\varepsilon^{0}(1-\varepsilon x)^{2(n-1)} & 0 &
\cdots\,\,\\
\;\; \vdots & \vdots & \vdots & \vdots & \ddots & \vdots\,\,
\end{pmatrix},
\end{equation*}
we obtain a fundamentally different general solution
$\tilde{\vy}$, which, for all
$x\in\mathbb{R}\backslash\{\frac{1}{\varepsilon}\}$, can {\it not}
be identified as a reparametrization of the integration constant
$\vc\mapsto\tilde{\vc}$ of the primary solution
\eqref{150401:1848} anymore:
\begin{equation}
\tilde{\vy}=\vG\!\left({\textstyle\frac{\tilde{x}}{1+\varepsilon
\tilde{x}}}\, ,\varepsilon\right)\cdot
\Big(e^{-\frac{\tilde{x}}{1+\varepsilon
\tilde{x}}\vA}\cdot\vc\Big)\neq e^{-\tilde{x}\vA}\cdot\tilde{\vc}.
\label{150407:1130}
\end{equation}
The fundamental difference between the solutions $\tilde{\vy}$
\eqref{150407:1130} and $\vy$ \eqref{150401:1848} already shows
itself in the fact that the former one
$\tilde{\vy}=\tilde{\vy}(\tilde{\vx},\varepsilon)$ has a {\it
permanent} non-removable singularity at $\tilde{x}=-1/\varepsilon$
independently of how $\vc$ is chosen, which thus implies that the
transformed solution $\tilde{\vy}$ has one {\it essential}
parameter more than the primary solution $\vy$, namely the group
parameter $\varepsilon$, which can {\it not} be generally absorbed
into the integration constant $\vc$.\footnote[2]{Note that
although the transformation for $x\mapsto \tilde{x}$ in
$\mathsf{T}_2$ \eqref{150323:2000} is identical to the one in
$\mathsf{L}_2$ \eqref{150403:1855} by also showing a non-removable
singularity at the inverse value of the group parameter, the {\it
full} transformation $\mathsf{T}_2$, however, does not induce this
singularity into the transformed general solution
\eqref{150323:2318}, simply because the corresponding
transformation $y\mapsto \tilde{y}$ in $\mathsf{T}_2$
\eqref{150323:2000} annihilates this singularity. Hence, in
contrast to transformation $\mathsf{L}_2$ \eqref{150403:1855}, the
group parameter in $\mathsf{T}_2$ \eqref{150323:2000} is {\it
non}-essential since it can be absorbed into the integration
constant to give the reparametrized general solution
\eqref{150323:2318}.}

Yet, $\tilde{\vy}$ \eqref{150407:1130} is not to
\eqref{150401:1848} the only {\it functionally different} general
solution which can be constructed by a symmetry transformation.
Infinitely many different general solutions can be obtained by
just relaxing the specification $\xi=x^2$ in $\mathsf{L}_2$
\eqref{150403:1420} and considering, for example, the more general
symmetry transformation
\begin{align}
\mathsf{L}_2^f: \;\;\; &
X=\xi(x,y_1,y_2,\dotsc)\,\partial_x+\sum_{n=1}^\infty\eta_n(x,y_1,y_2,\dotsc)\,\partial
_{y_n},\label{1504010:0846}\\
& \;\;\text{with}\;\;\xi=f(x),\;\;
\eta_n=y_n+\sum_{k=1}^{n-1}(-1)^{n-k}\binom{n-1}{k-1}\frac{d^{n-k}f(x)}{dx^{n-k}}y_{k+1},
\hspace{3.75cm}\nonumber
\end{align}
where $f$ is some arbitrary function. Hence, no unique and thus no
privileged general solution can be found for the infinite
hierarchy of differential equations \eqref{150401:1859}. As a
consequence, the infinite hierarchy \eqref{150401:1859} must be
identified as an unclosed and thus indeterminate set of equations,
irrespective of the fact that to every differential equation in
the hierarchy \eqref{150401:1859} one can {\it formally} associate
a solution function to it, which then, in a unique way, is coupled
to the next higher order equation.

And, once accepted that the hierarchy \eqref{150401:1830} is
unclosed, all invariant transformations which are admitted by this
system, as e.g. $\mathsf{L}_1$ \eqref{150403:1113}, $\mathsf{L}_2$
\eqref{150403:1420} and $\mathsf{L}^f_2$ \eqref{1504010:0846},
must then be identified not as symmetry transformations, but only
as weaker equivalence transformations which map between unclosed
systems
(see~e.g.~\cite{Ovsiannikov82,Meleshko96,Ibragimov04,Frewer14.1});
in this case they even constitute indeterminate transformations.
This identification is clearly supported when studying the most
general invariant transformation which the system
\eqref{150401:1830} can admit. It is given by two arbitrary
functions $f$ and $g$, one for the independent infinitesimal
$\xi=f(x,y_1)$ and one for the lowest order dependent
infinitesimal $\eta_1=g(x,y_1)$, which then both~uniquely assign
the functional structure for all remaining infinitesimals $\eta_n$
in the form $\eta_n=\eta_n(f(x,y_1),g(x,y_1),y_2,y_3,\dots,y_n)$,
for all $n\geq 2$.\footnote[3]{Note that each dependent
infinitesimal $\eta_n$ only shows a dependence up to order $n$ and
not beyond,~i.e.~it only depends on all dependent variables $y_m$
which appear below a considered level $n$, i.e. where $m\leq n$.}
This result shows complete arbitrariness in the choice for the
transformation of $x$ and $y_1$ to invariantly transform system
\eqref{150401:1830}, which, after all, is actually a trivial
result since the complete infinite hierarchy \eqref{150401:1859}
can also be equivalently written in the form of an underdetermined
solution as
\begin{equation}
y_{n+1}=(-1)^n\frac{d^n y_1}{dx^n},\;\; n=1,2,3,\dots,
\label{150415:1027}
\end{equation}
where it's more than obvious now that the considered system
\eqref{150401:1830} is not closed, since, through relation
\eqref{150415:1027}, all higher-order functions $y_{n+1}$ are
predetermined by the lowest order function $y_1$, but which itself
can be chosen completely arbitrarily. However, note that $y_1$ is
not privileged in the sense that only this function can be chosen
arbitrarily. Any function $y_{n^*}$ in the hierarchy
\eqref{150401:1859} can be chosen freely, where $n=n^*$ is some
arbitrary but fixed order in this hierarchy. Its\pagebreak[4]

\noindent underdetermined general solution can then be written as
\begin{equation}
\left.
\begin{aligned}
& y_{1} = (-1)^{n^*-1}\int y_{n^*}\, d^{n^*-1} x\\
&\;\vdots\\
& y_{n^*-k} = (-1)^k\int y_{n^*}\, d^k x\\
&\;\vdots\\
& y_{n^*-1} = -\int y_{n^*}\, dx\\
& y_{n^*+1} = - \frac{d y_{n^*}}{dx}\\
& y_{n^*+2} = \frac{d^2 y_{n^*}}{dx^2}\\
& \;\vdots\\
& y_{n^*+l} = (-1)^l\frac{d^l y_{n^*}}{dx^l}\\
& \;\vdots
\end{aligned}
~~~ \right \} \label{150411:1120}
\end{equation}
A corresponding invariance analysis certainly sees the same
effect, namely that {\it one} function, anywhere in the infinite
hierarchy \eqref{150401:1859}, can be chosen freely. Hence, since
through \eqref{150411:1120} any arbitrary general solution $\vy$
can be constructed, system \eqref{150401:1830} does not allow for
a {\it unique} general solution. The primary general solution
\eqref{150401:1848} is thus only one among an infinite set of
other, different possible general solutions which this system can
admit.

It should be noted here that our study only reveals the property
of global non-uniqueness when constructing a {\it general}
solution for an infinite system of differential equations which is
{\it unrestricted}$\,$; for example as for the plain system
\eqref{150401:1830} when no restrictions or any further conditions
on the solution manifold are imposed. In particular, our
statements do not invalidate the local uniqueness principle which
{\it may} exist for a system of ODEs once its restricted to
satisfy an initial condition.\footnote[2]{It is not exactly clear
yet in how far the well-defined local uniqueness principle for a
system of ODE initial value problems (Picard-Lindelöf theorem)
applies to systems which are infinite in dimension. Because, for
example, since for $\vA$~\eqref{150401:1830} not all matrix norms
are finite, they cannot be regarded as equivalent anymore. That
means, in order to guarantee the necessary Lipschitz continuity
for the function $\vA\cdot\vy$ on some interval, the infinite
matrix $\vA$ needs to satisfy the condition $\|\vA\|\leq L$ for
some finite Lipschitz constant $L$, thus leading to a conclusion
which now depends on the matrix norm used: For the maximum, row
and column norm, which give
$\|\vA\|_{\text{max}}=\|\vA\|_{\infty}=\|\vA\|_{1}=1$
respectively, the function $\vA\cdot\vy$ is Lipschitz continuous,
while for a norm which gives an infinite value, e.g. like the
Euclidean norm $\|\vA\|_{2}\rightarrow\infty$, the function
$\vA\cdot\vy$ is {\it not} Lipschitz continuous.} Independent of
whether this principle (Picard-Lindelöf theorem) uniquely applies
to infinite dimensional ODE initial value systems or not, for the
simple linear and homogeneous ODE structure \eqref{150401:1830},
however, it is straightforward to show that {\it local} uniqueness
in the solution for this particular infinite system {\it must}
exist, when specifying an initial condition at
$x=x_0\in\mathcal{I}$ inside some given local interval
$\mathcal{I}\subset\mathbb{R}$ (for the proof, see Appendix
\ref{C}). But, this local uniqueness interval $\mathcal{I}$ can be
quite narrow, and, depending on the chosen functions, can be even
of point-size only. In how narrow this local interval
$\mathcal{I}$ can be successively made is studied at a simple
example in Appendix \ref{C}.

Besides this, when specifying a particular initial condition, say
$\vy(x_0)=\vy_0$, or in component form $y_n(x_0)=y_{(0)n}$ for all
$n\geq 1$, then infinitely many and functionally independent
invariant (equivalence) transformations can be constructed which
all are compatible with this arbitrary but specifically chosen
initial condition. Because, since e.g.~the infinitesimals
$\xi=f(x,y_1)$ and $\eta_1=g(x,y_1)$ can be chosen arbitrarily,
one only has to guarantee that the initial condition
$y_n(x_0)=y_{(0)n}$, for all $n\geq 1$, gets mapped invariantly
into itself. This is achieved by demanding all infinitesimals to
satisfy the restrictions
\begin{align}
\xi(x,y_1)\Big\vert_{\{x=x_0;\vy=\vy_0\}}=0,\qquad
\eta_1(x,y_1)\Big\vert_{\{x=x_0;\vy=\vy_0\}}=0,\hspace{1cm}
\label{150513:0759}\\[0.75em]
\eta_n=\eta_n\Big(\xi(x,y_1),\eta_1(x,y_1),y_2,y_3,\dots,y_n\Big)
\Big\vert_{\{x=x_0;\vy=\vy_0\}}=0,\;\; n\geq 2,
\label{150513:0800}
\end{align}

\noindent where only the two infinitesimals $\xi$ and $\eta_1$ can
be chosen freely, while the remaining infinitesimals $\eta_n$, for
all $n\geq 2$, are predetermined differential functions of their
indicated arguments. The conditions \eqref{150513:0759}, in
accordance with \eqref{150513:0800}, can be easily fulfilled
e.g.~by restricting the arbitrary functions $\xi$ and $\eta_1$ to
\begin{equation}
\xi(x,y_1)=f_0(x,y_1)\cdot
e^{-\frac{\gamma_f^2}{(x-x_0)^2}},\qquad
\eta_1(x,y_1)=g_0(x,y_1)\cdot
e^{-\frac{\gamma_g^2}{(y_1-y_{(0)1})^2}}, \label{150513:0906}
\end{equation}
where $f_0$ and $g_0$ are again arbitrary functions, however, now
restricted to the class of functions which are increasing slower
than $e^{1/r^2}$ at $r=0$, where
$r=\sqrt{(x-x_0)^2/\gamma_f^2+(y_1-y_{(0)1})^2/\gamma_g^2}$. And,
since in this case all differential functions $\eta_n$, for $n\geq
2$, have the special non-shifted affine property
$\eta_n|_{\{\xi=0;\,\eta_1=0\}}=0$, the conditions
\eqref{150513:0800} all are automatically satisfied by the above
restriction \eqref{150513:0906}. Hence, an {\it infinite} set of
functionally independent (non-privileged) invariant solutions
$\vy=\vy(x)$ can be constructed from \eqref{150513:0906} which all
satisfy the given initial condition~$\vy(x_0)=\vy_0$.

\vspace{1em}\noindent {\it Remark on partial overlapping and
analytic continuation:}

\noindent Before closing this section let's briefly revisit the
result \eqref{150407:1130}. Important to mention here is that if
we expand the function $\frac{\tilde{x}}{1+\varepsilon \tilde{x}}$
into a power series around some arbitrary point $\tilde{x}=a\neq
-1/\varepsilon$, then the alternative general solution
$\tilde{\vy}$ \eqref{150407:1130} will map into a
reparametrization of the primary solution \eqref{150401:1848}, but
only in a very restrictive manner due to the existence of three
restrictions in order to ensure overall convergence (see
derivation \eqref{150409:1333} and \eqref{150409:2143}):
If,~in~$\mathbb{R}$, the chosen values for $\tilde{x}$,
$\varepsilon$ and $a$ satisfy the following three restrictions
simultaneously\footnote[2]{Note that since there are three
restrictions for three values, $\tilde{x}$, $\varepsilon$ and $a$,
they can not be chosen arbitrarily and independently anymore, i.e.
all three values depend on each other according to
\eqref{150409:2242}.~In general this combined set of restrictions
leads to a very narrow radius of convergence.}
\begin{equation}
\bigg\vert\,
\frac{(\tilde{x}-a)\,\varepsilon}{1+a\,\varepsilon}\,\bigg\vert <
1 ,\;\;\; \bigg\vert\,
\frac{a\,\varepsilon}{1+a\,\varepsilon}\,\bigg\vert < 1 ,\;\;
\text{and}\;\; \vert\, \tilde{x}\,\varepsilon\,\vert < 1,
\label{150409:2242}
\end{equation}
then, and only then, the symmetry transformation $\mathsf{L}_2$
\eqref{150407:1002} allows for a reparametrization of the primary
solution \eqref{150401:1848}
\begin{equation}
\tilde{\vy}=\vG\!\left({\textstyle\frac{\tilde{x}}{1+\varepsilon
\tilde{x}}}\, ,\varepsilon\right)\cdot
\Big(e^{-\frac{\tilde{x}}{1+\varepsilon
\tilde{x}}\vA}\cdot\vc\Big)= e^{-\tilde{x}\vA}\cdot\tilde{\vc},
\label{150407:1136}
\end{equation}
where the reparametrized integration constant $\tilde{\vc}$ is
then given by \eqref{150409:1325}, or equivalently
by~\eqref{150409:2319}, which both take the same form
\begin{equation}
\tilde{\vc}=\vG(0,\varepsilon)\cdot \vc. \label{150413:1036}
\end{equation}
However, for the remaining wide range of values  $\tilde{x}$,
$\varepsilon$ and $a$ within $\mathbb{R}$, namely for all those
values in which at least one of the three restrictions
\eqref{150409:2242} is violated, we still have, as was discussed
before, a second, fundamentally different general solution
$\tilde{\vy}$ \eqref{150407:1130} than as given by the primary
general solution $\vy$ \eqref{150401:1848}.

Further note that this particular example even allows for an
interesting {\it special case} when choosing $x_0=a=0$ as the
position, and $\vy_0=\tilde{\vc}$ \eqref{150413:1036} as the value
for an initial condition $\vy(x_0)=\vy_0$ of
system~\eqref{150401:1830}.~Because, since on the one side the
symmetry transformation \eqref{150407:1002} invariantly maps the
initial condition's space point from $x_0=0$ to $\tilde{x}_0=0$,
and since on the other side there exist a common domain
\eqref{150409:2242} in which the transformed general solution
\eqref{150407:1136} matches the primary general solution
\eqref{150401:1848}, the former (transformed) solution serves as
the functional continuation of the latter (primary) solution; but
only if, of course, both solutions originate from the same initial
condition, and if the primary solution \eqref{150401:1848}
converges on a given (untransformed) integration constant $\vc$.
To be explicit, we state that when considering the initial value
problem
\begin{equation}
\vy^\prime(x)=-\vA\cdot\vy(x),\;\;\text{with}\;\; \vy(0)=\vy_0,
\label{150519:1034}
\end{equation}
where $\vA$ is the infinite matrix given by \eqref{150401:1830}
and $\vy_0$ some arbitrary constant, then two solutions $\vy^A$
and $\vy^B$ exist, namely the primary solution $\vy^A$
\eqref{150401:1848} and the transformed solution $\vy^B$
\eqref{150407:1136}
\begin{align}
\vy^A(x) & = e^{-x\vA}\cdot\vy_0,\label{150413:1430}\\
\vy^B(x) & = \vG\!\left({\textstyle\frac{x}{1+\varepsilon x}}\,
,\varepsilon\right)\cdot \bigg[e^{-\frac{x}{1+\varepsilon
x}\vA}\cdot\left(\vG^{-1}(0,\varepsilon)\cdot\vy_0\right)\bigg],
\label{150413:1431}
\end{align}
which both satisfy the same initial condition
$\vy^A(0)=\vy^B(0)=\vy_0$, but where, according to the local
uniqueness principle for ODE initial value problems, each solution
serves as the functional continuation of the other solution on a
domain where it is not converging anymore. This domain depends on
the explicit initial value $\vy_0$ --- here $\vG$ is the infinite
group matrix \eqref{150407:1002} with its inverse
$\vG^{-1}(0,\varepsilon)=\vG(0,-\varepsilon)$ (see
\eqref{150519:0947}). In particular, if we choose the initial
value as given in \eqref{150413:1036}, i.e. if
$\vy_0=\tilde{\vc}$, with $\tilde{\vc}=\vG(0,\varepsilon)\cdot
\vc$, and where the primary (untransformed) integration constant
is~e.g.~fixed as $\vc\sim\boldsymbol{1}$, then the transformed
solution $\vy^B$ significantly extends the primary solution range
of $\vy^A$. The reason is that, since according to
\eqref{150409:2242} we are considering the special case $a=0$, the
primary solution $\vy^A$ \eqref{150413:1430} for this initial
condition only converges in the restricted domain $\vert
x\,\varepsilon\vert <1$, while the transformed solution $\vy^B$
\eqref{150413:1431} converges in the complete range
$x\in\mathbb{R}\backslash\{{\textstyle -\frac{1}{\varepsilon}}\}$.
Symbolically we thus have the relation $\vy^A\subset\vy^B$. For
more details and for a graphical illustration of these statements,
see Appendix \ref{F}. Surely, if we choose the initial value
$\vy_0$ such that on a specific domain the solution $\vy^B$ is not
converging, then $\vy^A$ serves as its continuation, i.e. we then
have the opposite relation $\vy^B\subset\vy^A$. See Table~\ref{T1}
in Appendix~\ref{F} for a collection of several choices in the
initial value $\vy_0$ and the corresponding domains for which the
solutions $\vy^A$ \eqref{150413:1430} and $\vy^B$
\eqref{150413:1431} are converging.

\subsection{Infinite nonlinear hierarchy of first order ODEs\label{3.2}}

In this section we will consider the nonlinear specification
$q_0=q_3=0,q_1=1/x,q_2=1/x^3$ of~\eqref{150401:1710}:
\begin{equation}
y_n^\prime-\frac{y_n}{x}=\frac{y_{n+1}^2}{x^3},\quad
n=1,2,3,\dotsc\label{150420:0806}
\end{equation}
This coupled system represents a genuine system of {\it nonlinear}
equations, since it cannot be reduced to a linear set of equations
as in the case of the corresponding single Riccati-ODE
\eqref{150323:1732} via the transformation $y(x)=x/z(x)$. In
addition, its underdeterminate solution cannot be written
compactly in closed form anymore as it was possible for the
previously considered linear system \eqref{150401:1859}, either
through \eqref{150415:1027}, or, more generally, through
\eqref{150411:1120}. But nevertheless, since \eqref{150420:0806}
is an {\it infinite forward} recurrence relation of {\it first}
order, i.e. where each term $y_{n+1}$ in the sequence depends on
the previous term $y_n$ in the functional form
$y_{n+1}=\mathcal{F}[y_n]=\pm\sqrt{x^3y_n^\prime-x^2y_n}$,\linebreak
it naturally acts again as an unclosed (underdetermined) system,
where, on any level in the prescribed hierarchy
\eqref{150420:0806}, exactly one function can be chosen freely.
This statement is again supported when performing an invariance
analysis upon system \eqref{150420:0806}, namely in the same way
as it was already discussed in the previous section: To transform
system \eqref{150420:0806} invariantly, complete arbitrariness
exists in that two arbitrary functions are available in order to
perform the transformation, one for the independent variable $x$
and one for any arbitrary but fixed chosen dependent variable
$y_{n^*}$, i.e.~where in effect the transformation of one function
$y_{n^*}=y_{n^*}(x)$ can be chosen absolutely freely. Hence, again
as in the previous case, infinitely many functionally independent
equivalence transformations can be constructed in the sense of
\eqref{150513:0906}, all being then compatible with any
specifically chosen initial condition $\vy(x_0)=\vy_0$.

As explained in the previous section in detail, an unclosed set of
differential equations, e.g. such as \eqref{150401:1859} or
\eqref{150420:0806}, does not allow for the construction of a {\it
unique general} solution that can cover {\it all} possible {\it
special} solutions these systems can admit. To explicitly
demonstrate this again for the nonlinear system
\eqref{150420:0806}, we first construct its most obvious general
solution based on the power series
\begin{equation}
y_n(x)=x^2\cdot \sum_{k=0}^\infty \lambda_{n,k}\, (x-a)^k,\;\;
n\geq 1, \label{150423:1555}
\end{equation}
where $a\in\mathbb{R}$ is some arbitrary expansion point. To be a
solution of \eqref{150420:0806}, the expansion coefficients have
to satisfy the following first order recurrence relation (see
Appendix \ref{G})
\begin{equation}
k\cdot \lambda_{n,k}+a\cdot (k+1)\cdot
\lambda_{n,k+1}+\lambda_{n,k} = \sum_{l=0}^k
\lambda_{n+1,k-l}\cdot \lambda_{n+1,l}\, ,\;\text{for all}\;\;
n\geq 1,\, k\geq 0. \label{150423:1603}
\end{equation}
For arbitrary but fixed given initial values $c_n:=\lambda_{n,0}$,
this recurrence relation can be explicitly solved for all higher
orders relative to the expansion index $k\geq 1$. If $a\neq 0$,
the first four expansion coefficients, for all $n\geq 1$, are then
given as
\begin{equation}
\left .
\begin{aligned}
\lambda_{n,0} & = c_n,\\[0.5em]
\lambda_{n,1} & = -\frac{1}{a}\cdot
\Big(c_n-c_{n+1}^2\Big),\\[0.5em]
\lambda_{n,2} & =
\frac{1}{a^2}\cdot\Big(c_n-2c_{n+1}^2+c_{n+1}\cdot c_{n+2}^2\Big),\\[0.5em]
\lambda_{n,3} & =
-\frac{1}{3a^3}\cdot\Big(3c_n-9c_{n+1}^2+9c_{n+1}\cdot
c_{n+2}^2-2c_{n+1}\cdot c_{n+2}\cdot
c_{n+3}^2-c_{n+2}^4\Big),\;\\[0.25em]
& \;\;\vdots
\end{aligned}
~~~~~~~~ \right\}\!\!\!\!\!\! \label{150423:1617}
\end{equation}

\noindent while if $a=0$, they are given as\footnote[2]{For $a=0$
the recurrence relation \eqref{150423:1603} can be solved by
making use for example of a generating function or the more
general $Z$-transform (see e.g. \cite{Zeidler04}). Hereby should
be noted that relation \eqref{150423:1603} is a 1-dimensional
recurrence relation of first order for any arbitrary but fixed
order $n$.}
\begin{equation}
\left .
\begin{aligned}
\negthickspace \negthickspace\negthickspace
\phantom{c_n =} \lambda_{n,0} & = e^{2^{1-n}\cdot\, \sigma_1}=c_n,\\[0.5em]
\lambda_{n,1} & = e^{(2^{1-n}-1)\cdot\sigma_1}\cdot
\sigma_2,\\[0.5em]
\lambda_{n,2} & = 2^{-n}\cdot
e^{(2^{1-n}-2)\cdot\sigma_1}\cdot\left(\sigma_2^2\cdot
(2^n-3^n)+2\cdot
3^{-1+n}\cdot e^{\sigma_1}\cdot\sigma_3\right),\\[0.5em]
\lambda_{n,3} & = 2^{-1-n}\cdot
e^{(2^{1-n}-3)\cdot\sigma_1}\cdot\Big(\sigma_2^3\cdot
(2^{1+n}+2^{2+2n}-2\cdot 3^{1+n})\\
&\qquad\qquad\qquad\qquad\qquad\quad\;\; +(4\cdot
3^n-4^{1+n})\cdot e^{\sigma_1}\cdot \sigma_2\cdot\sigma_3+4^n\cdot
e^{2\sigma_1}\cdot\sigma_4\Big),\\[-0.5em]
& \;\;\vdots
\end{aligned}
~~ \right\} \label{150423:1618}
\end{equation}

\noindent where $\vsigma=(\sigma_1,\sigma_2,\dotsc,
\sigma_n,\dotsc)$ is a new infinite set of integration constants,
which now, instead of $\vc=(c_1,c_2,\dotsc,c_n,\dotsc)$, take the
place for the freely selectable parameters in the general solution
\eqref{150423:1555} as soon as the expansion point $a$ turns to
zero; simply because in this singular case all constants $c_n$
within \eqref{150423:1603} cannot be chosen independently anymore
as they would all depend on the single choice of the first
parameter $\sigma_1$ as given above in the first line of
\eqref{150423:1618}.

That solution \eqref{150423:1555}, along with either
\eqref{150423:1617} or \eqref{150423:1618}, represents a general
solution is obvious, since to each solution $y_n$ of the {\it
first} order system \eqref{150420:0806} one can associate to it a
free parameter for all $n\geq 1$, either $c_n$ if $a\neq 0$, or
$\sigma_n$ if $a=0$. But this {\it general} solution is not unique
as it does not cover all possible {\it special} solutions which
that system \eqref{150420:0806} can admit. For example, if we
consider the following independent\footnote[2]{The only dependence
which exists between these three different solutions is that
$\lim_{n\to\infty}y_n^{(3)}=y_{n^*}^{(1)}=x^2$~for all $n^*\geq
1$.} special solutions of \eqref{150420:0806}
\begin{align}
y_n^{(1)}(x) & =  x^2, \;\;\text{for all}\;\; n\geq 1,\label{150423:1914}\\[0.5em]
y_n^{(2)}(x) & = \begin{cases}\, -1, \;\;\text{for}\;\;
n=1,\\
\;\;\; x, \;\;\text{for}\;\; n=2,\\
\;\;\:\, 0, \;\;\text{for all}\;\; n\geq 3, \end{cases}\label{150423:1946}\\[0.5em]
y_n^{(3)}(x) & = \frac{1}{5^{2^{-n}}}\left[\,\prod_{k=0}^{n-1}
\left(1+\frac{1}{2^{k-2}}\right)^{2^{k-n}}\,\right]\cdot
x^{2+\frac{1}{2^{n-2}}}, \;\;\text{for all}\;\; n\geq
1,\label{150423:2018}
\end{align}

\noindent only solution $y_n^{(1)}$ is covered by the general
solution \eqref{150423:1555} for all $n\geq 1$, in choosing either
$c_n=1$ (if $a\neq 0$), or $\sigma_n=0$ (if $a=0$). However, the
special solution $y_n^{(2)}$, and in general also
$y_n^{(3)}$,\linebreak are not covered. The reason for $y_n^{(2)}$
\eqref{150423:1946} is obvious, because it's a polynomial with a
smaller degree than $y_n$ \eqref{150423:1555}, which itself is
always at {\it least} of second order for all $n\geq 1$.~And to
see the reason for $y_n^{(3)}$ \eqref{150423:2018}, it's
sufficient to explicitly evolve it up to third order in $n$:
\begin{equation}
y_1^{(3)}=x^4,\;\;\;\; y_2^{(3)}=\sqrt{3}\cdot x^3,\;\;\;\;
y_3^{(3)}=\sqrt{2}\cdot\sqrt[4]{3}\cdot x^{5/2},\;\;\;\cdots
\end{equation}
Because, if $a=0$, one has to choose  $\sigma_1\rightarrow
-\infty$ in order to obtain the lowest order particular solution
$y_1^{(3)}$, which then turns into a contradiction when trying to
determine $\sigma_2$ for the next higher order particular solution
$y_2^{(3)}$. This indeterminacy will then propagate through all
remaining orders $n\geq 3$, i.e. for $a=0$ no consistent set of
expansion coefficients $\lambda_{n,k}$~\eqref{150423:1618} can be
determined to generate the special solution \eqref{150423:2018}
from the general solution \eqref{150423:1555}. If, however, the
expansion point of the general solution \eqref{150423:1555} is
chosen to be $a\neq 0$, then one obtains a coverage for the
special solution $y_n^{(3)}$ \eqref{150423:2018}, but only
partially, namely only in the domain $|x-a|<|a|$ (for the proof,
see Appendix \ref{H}). Hence, the general solution
\eqref{150423:1555} is not unique, since other general solutions
of \eqref{150420:0806} must exist in order to completely cover for
example the special solutions $y_n^{(2)}$ \eqref{150423:1946} and
$y_n^{(3)}$ \eqref{150423:2018} for all $x\in\mathbb{R}$.

To close this investigation it is worthwhile to mention that the
infinite nonlinear hierarchy of {\it coupled} equations
\eqref{150420:0806} allows for two invariant Lie group actions
which are {\it uncoupled}. For all $n\geq 1$, these~are:
\begin{equation}
\left. \begin{aligned} \mathsf{T}_1^\infty: & \;\;\;
\tilde{x}=e^{\varepsilon_1} x,\;\;\;
\tilde{y}_n=e^{2\varepsilon_1}y_n,\\[0.75em]
\mathsf{T}_2^\infty: & \;\;\; \tilde{x}=
\frac{x}{1-\varepsilon_2\, x},\;\;\;
\tilde{y}_n=\frac{y_n}{1-\varepsilon_2\, x},
\end{aligned}
~~~ \right \} \label{150424:2033}
\end{equation}

\noindent being the equivalent invariances to the scaling symmetry
$\mathsf{T}_1$ and projective symmetry $\mathsf{T}_2$ of the
corresponding {\it single} Riccati-ODE \eqref{150323:1732} in
\eqref{150323:2000} respectively. Their infinitesimal form has the
structure
\begin{equation}
\left. \begin{aligned} \mathsf{T}_1^\infty: & \;\;\; X_1^\infty=\,
x\partial_x+2y_1\partial_{y_1}+2y_2\partial_{y_2}+\cdots
+2y_n\partial_{y_n}+\cdots,\\[0.75em]
\mathsf{T}_2^\infty: & \;\;\; X_2^\infty=\,
x^2\partial_x+xy_1\partial_{y_1}+xy_2\partial_{y_2}+\cdots
+xy_n\partial_{y_n}+\cdots,
\end{aligned}
~~~ \right \} \label{150424:2258}
\end{equation}

\noindent and, as proven in Appendix \ref{J}, they are the two
only possible {\it non-coupled} Lie-point invariances which the
coupled hierarchy \eqref{150420:0806} of first order Riccati-ODEs
can admit. As was demonstrated in the previous section, these
invariant (equivalence) transformations \eqref{150424:2033} can
now be used to either generate new {\it additional} special
solutions or to generate functionally {\it different} general
solutions by just transforming \eqref{150423:1555} respectively.
Hereby note that the special solution $y_n^{(1)}$
\eqref{150423:1914} is an {\it invariant solution} with respect to
the scaling invariance $\mathsf{T}_1^\infty$ \eqref{150424:2033},
which even can be prolonged to the more general invariant solution
\begin{equation}
y_n^{(1)}(x;\tau)=e^{\tau\cdot2^{1-n}}\cdot x^2,
\end{equation}
which then involves a free parameter $\tau\in\mathbb{R}$ for all
$n\geq 1$.

\section{Conclusion\label{4}}

At the example of first order ODEs this study has shown that an
infinite and forward recursive hierarchy of differential equations
carries all features of an unclosed system, and that,
conclusively, all admitted invariance transformations must be
identified as equivalence transformations only. To obtain from
such systems an invariant solution which shows a certain
particular functional structure is ultimately without value, since
infinitely many functionally different and non-privileged
invariant solutions can be constructed, even if sufficient initial
conditions are additionally imposed. In order to obtain valuable
results, the infinite system needs to be closed by posing
modelling assumptions which have to reflect the structure of the
underlying (higher abstraction level) equations from which the
infinite system~emerges.

It is clear that this insight is not restricted to ODEs, but that
it holds for differential equations of any type as soon as the
infinite hierarchy is of a forward recursive nature. For example,
as it's the case for the infinite Friedmann-Keller hierarchy of
PDEs for the multi-point moments in statistical turbulence theory.
As it is discussed in detail in \cite{Frewer14.1}\linebreak and
further in \cite{Frewer15.1}, this infinite system is undoubtedly
unclosed and that it's simply\linebreak without any value,
therefore, to determine particular invariant solutions if no prior
modelling\linebreak assumptions are invoked on that system.

\appendix
\titleformat{\subsection}
{\normalfont\itshape\filcenter}{\normalfont\thetitle.}{0.5em}{}

\section{Infinite backward versus infinite forward differential
recurrence relations\label{A}}

\subsection{Example for an infinite backward differential recurrence relation\\
(Closed system with unique solution manifold)\label{A1}}

Let us consider the Cauchy problem \eqref{150430:1313} of
Example~\hyperref[E5]{5} in the simplified form $a=1$ and $b=c=0$,
along with an initial function $\phi$ which is normalized to
$\int_{-\infty}^\infty \phi(x)dx=1$. Then this initial value
problem~\eqref{150430:1313} has the {\it unique} solution
\begin{equation}
u(t,x) = \frac{1}{\sqrt{4\pi t}}\int_{-\infty}^\infty
e^{-\frac{(x-x^\prime)^2}{4t}}\phi(x^\prime) \,
dx^\prime,\;\;\text{for}\;\; t\geq 0, \label{150501:1915}
\end{equation}
which by construction, due to $c=0$, automatically satisfies the
normalization constraint \eqref{150501:1337}. To study this
uniqueness issue on the corresponding moment induced ODE system
\eqref{150501:1415}, let us\linebreak first consider the {\it
unrestricted} system
\begin{equation}
\frac{d u_n}{dt} = n\cdot(n-1)\cdot u_{n-2},\;\; n\geq
0,\label{150504:1334}
\end{equation}
which, as an infinite {\it backward} recursive system, can be
written into the equivalent form
\begin{equation}
\left .
\begin{aligned}
& \frac{du_0(t)}{dt}=0,\quad \,\,\frac{d^{n+1}
u_{2n+2}(t)}{dt^{n+1}}
=(2n+2)!\cdot u_0(t),\;\; n\geq 0,\\[0.5em]
\text{and} \;\;\, &\frac{du_1(t)}{dt}=0,\quad \frac{d^m
u_{2m+1}(t)}{dt^m}=(2m+1)!\cdot u_1(t),\;\; m\geq 1.
\end{aligned}
~~~~ \right\} \label{150504:1341}
\end{equation}

\noindent This system can then be uniquely integrated to give the
{\it general} solution
\begin{equation}
\!\!\!\!\!\!\left .
\begin{aligned}
u_0(t)& =c_0,\\[0.5em]
u_{2n+2}(t)
&=(2n+2)!\int_0^{t_n=t}\int_0^{t_{n-1}}\!\cdots\int_0^{t_0}u_0(t^\prime)\,
dt^\prime\, dt_0\cdots
dt_{n-1}+\sum_{k=0}^{n}\frac{q^{(1)}_{n,k}}{k!}\, t^k,\; n\geq 0,\\[0.5em]
\quad u_1(t)&=c_1,\\[0.5em]
u_{2m+1}(t)&=(2m+1)!\int_0^{t_m=t}\int_0^{t_{m-1}}\!\cdots\int_0^{t_1}u_1(t^\prime)\,
dt^\prime\, dt_1\cdots
dt_{m-1}+\sum_{k=0}^{m-1}\frac{q^{(2)}_{m,k}}{k!}\, t^k,\; m\geq
1,
\end{aligned}
\right\}\label{150501:1916}
\end{equation}

\noindent with the expansion coefficients given as
\begin{equation}
\left .
\begin{aligned}
q^{(1)}_{n,k}&=\frac{(2n+2)!}{(2n+2-2k)!}\, c_{2n+2-2k},\;\; n\geq
0;\;\; 0\leq k\leq n,\\[0.5em]
q^{(2)}_{m,k}&=\frac{(2m+1)!}{(2m+1-2k)!}\, c_{2m+1-2k},\;\; m\geq
1;\;\; 0\leq k\leq m-1,
\end{aligned}
~~~ \right\}
\end{equation}

\noindent where all $c_n$ for $n\geq 0$ are arbitrary integration
constants. Hence we see that the unrestricted system
\eqref{150504:1334} provides a general solution
\eqref{150501:1916} which only involves arbitrary constants, i.e.
the unrestricted system \eqref{150504:1334} provides a {\it
unique} general solution. Because, when restricting this system to
the underlying PDE's initial condition $u(0,x)=\phi(x)$, with
$\int_{-\infty}^\infty\phi(x)dx=1$, which for the ODE system
\eqref{150504:1334} takes the form
\begin{equation}
u_n(0)=\int_{-\infty}^\infty x^n\cdot \phi(x)\, dx,\;\; n\geq
0,\;\;\text{with}\;\; u_0(0)=1, \label{150501:1934}
\end{equation}
it will turn the general solution \eqref{150501:1916} into a
unique and fully determined solution, where the integration
constants are then given by
\begin{equation}
c_n=u_n(0), \;\; n\geq 0,\;\;\text{with}\;\; c_0=1.
\end{equation}

\subsection{Example for an infinite forward differential recurrence relation\\
(Unclosed system with non-unique solution manifold)\label{A2}}

Now, let's consider the case $a=1$, $b=0$ and $c=-1$, where again
the initial condition function $\phi$ is normalized to
$\int_{-\infty}^\infty \phi(x)dx=1$. To solve this initial value
problem (Cauchy problem)
\begin{equation}
\partial_t u=\partial_x^2 u-x^2 u,\;\:\text{for}\;\: t\geq
0,\;\;\text{with}\;\; u(0,x)=\phi(x), \label{150504:2203}
\end{equation}
it is necessary to realize that the following nonlinear point
transformation \citep{Polyanin02}\footnote[2]{This continuous
point transformation is {\it not} a group transformation, as it
neither includes a group parameter nor does it include the unique
continuously connected identity transformation from which any
infinitesimal mapping can emanate.}
\begin{equation}
\tilde{t}=\frac{1}{4}\cdot\left(e^{4t}-1\right),\quad
\tilde{x}=x\cdot e^{2t},\quad \tilde{u}=u\cdot
e^{-\frac{1}{2}x^2-t}, \label{150504:2314}
\end{equation}
which has the unique inverse transformation
\begin{equation}
t=\frac{1}{4}\ln\big(1+4\tilde{t}\hspace{.06cm}\big),\quad
x=\frac{\tilde{x}}{\sqrt{1^{\vphantom{A^A}}+4\tilde{t}}},\quad
u=\tilde{u}\cdot \sqrt[4]{1+4\tilde{t}}\cdot
e^{\frac{1}{2}\cdot\frac{\tilde{x}^2}{1^{\vphantom{1}}+4\tilde{t}}},
\end{equation}
maps the original Cauchy problem \eqref{150504:2203} into the
following Cauchy problem for the standard diffusion equation with
constant coefficients:
\begin{equation}
\partial_{\tilde{t}\,}
\tilde{u}=\partial_{\tilde{x}\,}^2\tilde{u},\;\:\text{for}\;\:
\tilde{t}\geq 0,\;\;\text{with}\;\;
\tilde{u}(0,\tilde{x})=\phi(\tilde{x})\cdot
e^{-\frac{1}{2}\tilde{x}^2}. \label{150504:2210}
\end{equation}
Important to note here is that the initial time $t=0$ as well as
the relevant time range $t\in [0,\infty)$ both get invariantly
mapped to $\tilde{t}=0$ and $\tilde{t}\in [0,\infty)$
respectively. Hence, the unique solution of the transformed Cauchy
problem \eqref{150504:2210} is thus again given by
\eqref{150501:1915}, but now in the form
\begin{equation}
\tilde{u}(\tilde{t},\tilde{x}) =
\frac{1}{\sqrt{4\pi^{\vphantom{A^1}}
\tilde{t}}}\int_{-\infty}^\infty
e^{-\frac{(\tilde{x}-\tilde{x}^\prime)^2}{4^{\vphantom{1}}\tilde{t}}}\phi(\tilde{x}^\prime)
\, e^{-\frac{1}{2}\tilde{x}^{\prime\, 2}} \,
d\tilde{x}^\prime,\;\;\text{for}\;\; \tilde{t}\geq 0,
\label{150504:2302}
\end{equation}
which then, according to transformation \eqref{150504:2314}, leads
to the unique solution for the original Cauchy
problem~\eqref{150504:2203}
\begin{equation}
u(t,x)=
\frac{e^{\frac{1}{2}x^2+t}}{\sqrt{\pi\big(e^{4t}-1^{\vphantom{1}}\big)}}
\int_{-\infty}^\infty
e^{-\frac{(e^{2t}x-x^\prime)^2}{e^{4t^{\vphantom{:}}}-1}}\phi(x^\prime)
\,\, e^{-\frac{1}{2}x^{\prime\, 2}} \,
dx^\prime,\;\;\text{for}\;\; t\geq 0. \label{150504:2344}
\end{equation}
Considering, however, the corresponding moment induced infinite
ODE system \eqref{150501:1415} for \eqref{150504:2203}, we will
now show that this system is {\it not} uniquely specified and thus
has to be identified as an unclosed system even if sufficient
initial conditions are imposed. In clear contrast to its
associated higher level PDE system \eqref{150504:2203}, which, as
a Cauchy problem, is well-posed by providing the unique solution
\eqref{150504:2344}. To see this, let us first again consider the
{\it unrestricted}~ODE~system
\begin{equation}
\frac{du_n}{dt} = n\cdot (n-1)\cdot u_{n-2}-u_{n+2},\;\; n\geq 0,
\label{150507:1032}
\end{equation}
which can be rewritten into the equivalent and already solved
form\footnote[2]{In the following we agree on the definitions that
$\frac{d^0}{dt^{0}}=1$, $\frac{d^{q<0}}{dt^{q}}=0$, and
$\sum_{i=0}^{q<0}=0$.}
\begin{equation}
\left.
\begin{aligned}
u_{2n+2}(t)&=(-1)^{n+1}\sum_{i=0}^\infty
A^{(1)}_i(n)\frac{d^{n+1-2i}}{dt^{n+1-2i}}\, u_0(t),\;\; n\geq
0,\\
u_{2m+1}(t)&=(-1)^{m}\sum_{j=1}^\infty
A^{(2)}_j(m)\frac{d^{m+2-2j}}{dt^{m+2-2j}}\, u_1(t),\;\; m\geq 1,
\end{aligned}
~~~ \right\}\label{150507:1030}
\end{equation}
where the coefficients $A^{(1)}_i(n)$ and $A^{(2)}_j(m)$ are
recursively defined as:
\begin{equation}
\left .\!\!\!\!\!\!
\begin{aligned}
\bullet &\;\:\text{{\it Initial seed for $A^{(1)}_i(n)$}:}\;\;\;
A^{(1)}_0(-1)=1,\;\text{and}\;\: A^{(1)}_0(n)=1,\;\text{for all $n\geq0$},\hspace{1.4cm}\\[0.5em]
&\;\;\: A^{(1)}_i(n) = \sum_{k=0}^{n-(2i-1)} (2n-2k)\cdot
(2n-1-2k)\cdot
A^{(1)}_{i-1}(n-2-k),\;\;\; i\geq 1,\;\; n\geq 0,\hspace{0.8cm}\\[0.5em]
\bullet &\;\:\text{{\it Initial seed for $A^{(2)}_j(m)$}:}\;\;\;
A^{(2)}_1(0)=1,\;\text{and}\;\:
A^{(2)}_1(m)=1,\;\text{for all $m\geq 1$},\hspace{1.375cm}\\[0.5em]
&\;\;\: A^{(2)}_j(m) = \sum_{k=1}^{m-(2j-3)} (2m-2k)\cdot
(2m+1-2k)\cdot A^{(2)}_{j-1}(m-1-k),\;\;\; j\geq 2,\;\; m\geq 1.
\end{aligned}
\right\}
\end{equation}

\noindent In contrast to the {\it general} solution
\eqref{150501:1916} of the previously considered {\it
unrestricted} system \eqref{150504:1334}, we see that the degree
of underdeterminedness in the above determined {\it general}
solution \eqref{150507:1030} is fundamentally different and higher
than in \eqref{150501:1916}.~Instead of integration constants
$c_n$, we now have two integration functions $u_0(t)$ and $u_1(t)$
which can be chosen freely. Their (arbitrary) specification will
then determine all other solutions for $n\geq 0$ and $m\geq 1$
according to \eqref{150507:1030}. The reason for having two free
functions and not infinitely many free constants is that system
\eqref{150507:1032} defines a {\it forward} recurrence relation
(of order two)\footnote[2]{The order of the recurrence relation is
defined relative to the differential operator.} that needs {\it
not} to be integrated in order to determine its general solution,
while system \eqref{150504:1334}, in contrast, defines a {\it
backward} recurrence relation (of order two) which needs to be
integrated to yield its general solution.

To explicitly demonstrate that \eqref{150507:1030} is not a {\it
unique} general solution, we have to impose the corresponding
initial condition $u(0,x)=\phi(x)$, with $\int_{-\infty}^\infty
\phi(x)\, dx =1$, which led to the unique solution
\eqref{150504:2344} of the underlying PDE system
\eqref{150504:2203}. For the current ODE system
\eqref{150507:1032} this condition takes again the same form
\eqref{150501:1934} as it did for previous ODE system
\eqref{150504:1334}:
\begin{equation}
u_n(0)=\int_{-\infty}^\infty x^n\cdot \phi(x)\, dx,\;\; n\geq
0,\;\;\text{with}\;\; u_0(0)=1. \label{1505017:1438}
\end{equation}
The easiest way to perform this implementation is to choose the
two arbitrary functions $u_0(t)$ and $u_1(t)$ as analytical
functions which can be expanded as power series
\begin{equation}
u_0(t)=\sum_{k=0}^\infty \frac{c^{(1)}_k}{k!}\, t^k,\quad\;
u_1(t)=\sum_{k=0}^\infty \frac{c^{(2)}_k}{k!}\, t^k,
\label{150508:0818}
\end{equation}
where $c^{(1)}_k$ and $c^{(2)}_k$ are two different infinite sets
of constant expansion coefficients. By inserting this Ansatz into
the general solution \eqref{150507:1030} and imposing the initial
conditions \eqref{1505017:1438} will then uniquely specify these
coefficients in a recursive manner as
\begin{equation}
\left .
\begin{aligned}
c^{(1)}_k =0,\;\; k<0; \quad\;\; c^{(1)}_k & =(-1)^k\cdot
u_{2k}(0)-\sum_{i=1}^\infty
A^{(1)}_i(k-1)\cdot c^{(1)}_{k-2i},\;\;k\geq 0,\\[0.5em]
c^{(2)}_k =0,\;\; k<0; \quad\;\; c^{(2)}_k & =(-1)^k\cdot
u_{2k+1}(0)-\sum_{i=1}^\infty A^{(2)}_{i+1}(k)\cdot
c^{(2)}_{k-2i},\;\;k\geq 0.
\end{aligned}
~~~ \right\}\label{150507:2005}
\end{equation}

\noindent Indeed, the two solutions \eqref{150508:0818} with the
above determined coefficients \eqref{150507:2005} form the
analytical part of the corresponding unique PDE moment solutions
relative to \eqref{150504:2344}. For example, choosing the
non-symmetric and to one normalized initial condition function
$\phi(x)=\frac{1}{\sqrt{\pi}}e^{-(x-1)^2}$ will give the first two
unique PDE moment solutions as
\begin{equation}
\left .
\begin{aligned}
u_0(t)&=\int_{-\infty}^\infty x^0\cdot u(t,x)\, dx = \frac{2\,
e^{-\frac{1}{3}\left(1-\frac{4}{1+3e^{4t^{\vphantom{.}}}}\right)+t}}
{\sqrt{1+3e^{4t^{\vphantom{1}}}}},\;\;
t\geq 0,\\[0.75em]
u_1(t)&=\int_{-\infty}^\infty x^1\cdot u(t,x)\, dx =\frac{8\,
e^{-\frac{1}{3}\left(1-\frac{4}{1+3e^{4t^{\vphantom{.}}}}\right)+3t}}
{\sqrt{\left(1+3e^{4t}\right)^3}},\;\;
t\geq 0,
\end{aligned}
~~~ \right\}\label{150508:0825}
\end{equation}

\noindent and, if these were Taylor expanded around $t=0$, they
would exactly yield the first two power series solutions
\eqref{150508:0818} of the associated infinite ODE system
\eqref{150507:1032}. But, the Taylor expansions of both functions
\eqref{150508:0825} only converge in the limited range $0\leq t <
\frac{1}{4}\sqrt{\pi^{2^{\vphantom{.}}}+(\ln 3)^2}\sim 0.83$.

That means, our initial assumption that the first two ODE
solutions $u_0(t)$ and $u_1(t)$ are analytical functions on the
global and unlimited scale $t\in[0,\infty)$ is thus not correct.
Only for a very limited range this functional choice
\eqref{150508:0818} is valid. But, if we don't know the full scale
PDE solutions \eqref{150508:0825} beforehand, how then to choose
these two unknown functions $u_0(t)$ and $u_1(t)$ for the infinite
ODE system \eqref{150507:1030}? The clear answer is that there is
no way without invoking a prior modelling assumption on the ODE
system itself. Even if we would choose specific functions $f_0(t)$
and $f_1(t)$, which for $u_0(t)$ and $u_1(t)$ are valid on any
larger scale than the limited analytical Ansatz
\eqref{150508:0818}, we still have the problem that this
particular solution choice is not unique, because one can always
add to this choice certain independent functions which give no
contributions when evaluated at the initial point $t=0$. For
example, if $u_0(t)=f_0(t)$ and $u_1(t)=f_1(t)$, and if both
functions $f_0$ and $f_1$ satisfy the given initial conditions at
$t=0$,~then
\begin{equation}
u_0(t)=f_0(t)+ \psi_0(t)\cdot e^{-\frac{\gamma_0^2}{t^2}},\qquad
u_1(t)=f_1(t)+ \psi_1(t)\cdot e^{-\frac{\gamma_1^2}{t^2}},
\label{150508:1335}
\end{equation}
is also a possible solution choice which satisfies the same
initial conditions, where $\psi_0(t)$ and~$\psi_1(t)$ are again
arbitrary functions, with the only restriction that, at the
initial point $t=0$, they have to increase slower than
$e^{\gamma_0^2/t^2}$ and $e^{\gamma_1^2/t^2}$ respectively.

That no unique solution can be constructed a priori provides the
reason that the PDE induced ODE system \eqref{150507:1030},
although infinite in dimension, has to be treated as an unclosed
system. It involves more unknown functions than there are
determining equations, although {\it formally}, in a bijective
manner, to each function within the hierarchy a corresponding
equation can be mapped to. But, since the hierarchy
\eqref{150507:1032} can be equivalently rewritten into the form
\eqref{150507:1030}, it explicitly reveals the fact that exactly
two functions $u_0(t)$ and $u_1(t)$ in this hierarchy remain
unknown, and without the precise knowledge of their global
functional structure all remaining solutions $u_n(t)$ for $n\geq
2$ then remain unknown too. And, since the equivalently rewritten
form \eqref{150507:1030} already represents the {\it general}
solution of the original infinite ODE system \eqref{150507:1032},
the general solution itself is unclosed as well. In other words,
the general solution \eqref{150507:1030} is not a {\it unique}
general solution. The degree of arbitrariness in having two
unknown functions cannot be reduced, even when imposing initial
conditions, simply due to the existing modus operandi in the sense
of \eqref{150508:1335} when constructing possible valid solutions.

Hence, posing any initial conditions are thus not sufficient to
yield a unique solution for the (lower abstraction level) ODE
system \eqref{150507:1030} as they are for the (higher abstraction
level) PDE equation \eqref{150504:2203}. Without a prior modelling
assumption on the ODE system \eqref{150507:1030}, this system
remains unclosed. Fortunately, the solutions of this particular
case \eqref{150508:0825} possessed an analytical part in their
functions for which the assumed Ansatz \eqref{150508:0818}
expressed the correct functional behavior, though only in a very
narrow and limited range. But, of course, for more general cases
such a partial analytical structure is not always necessarily
provided, and an Ansatz as \eqref{150508:0818} would then be
misleading.

\section{Alternative method in constructing a global
transformation\label{B}}

Besides Lie's central theorem, a more efficient way to determine
the global 1-parametric symmetry transformation of $y_n$ from its
infinitesimal form \eqref{150403:1420} for $n\geq 2$ is, in this
particular case, to make use of the underlying recurrence relation
\eqref{150401:1859} along with the transformation rule
\eqref{150404:1245} in its more general form:
\begin{align}
\tilde{y}_n &
=-\tilde{y}_{n-1}^\prime=-\frac{d\tilde{y}_{n-1}}{d\tilde{x}}
=-\frac{d\tilde{y}_{n-1}(x,y_1,y_2,\dotsc,y_{n-1})}{d\tilde{x}(x)}
=-\left(\frac{\partial\tilde{x}}{\partial x}\right)^{-1}
\left(\,\sum_{q=1}^{n-1}\frac{\partial \tilde{y}_{n-1}}{\partial
y_{q}}y^\prime_{q}+\frac{\partial \tilde{y}_{n-1}}{\partial
x}\right)\nonumber
\end{align}
\begin{align}
\tilde{y}_n & =\left(\frac{\partial\tilde{x}}{\partial
x}\right)^{-1} \left(\,\sum_{q=1}^{n-1}\frac{\partial
\tilde{y}_{n-1}}{\partial y_{q}}y_{q+1}-\frac{\partial
\tilde{y}_{n-1}}{\partial x}\right) = (1-\varepsilon
x)^2\left(\,\sum_{q=1}^{n-1}\frac{\partial
\tilde{y}_{n-1}}{\partial y_{q}}y_{q+1}-\frac{\partial
\tilde{y}_{n-1}}{\partial x}\right)\nonumber\qquad\qquad\\
& = \, \sum_{k=1}^{n-1} B_{n,k}\,\varepsilon^{n-k-1}(1-\varepsilon
x)^{n+k-1}\, e^{\varepsilon}\, y_{k+1} ,\;\;\text{for all}\;\;
n\geq 2, \label{150404:1253}
\end{align}

\noindent where the coefficients $B_{n,k}$ are defined via the
following 2-dimensional recurrence relation\footnote[2]{In order
to obtain from a (1+1)-dimensional recurrence relation a unique
solution it has to be supplemented by one initial condition and
two zero-dimensional boundary conditions; in full analogy to the
situation for PDEs.}
\begin{equation}
B_{n,k}=(n-2+k)\cdot B_{n-1,k}+B_{n-1,k-1},\;\; \text{for $n\geq
3$ and $k=1,2,\dotsc,n-1$}, \label{150518:1452}
\end{equation}
with the initial condition $B_{2,1}=1$, and the boundary
conditions $B_{n,0}=0$ (left boundary) and $B_{n,n}=0$ (right
boundary) for all $n\geq 2$.

\section{Local uniqueness proof and an example on its range\label{C}}

\noindent {\it Proposition:} Given is the following initial value
problem for the infinite ODE system \eqref{150401:1830}
\begin{equation}
\vy^\prime=-\vA\cdot\vy,\;\;\text{with}\;\; \vy(x_0)=\vy_0,
\label{150417:0935}
\end{equation}
in some local interval $\mathcal{I}\subset\mathbb{R}$, where
$x_0\in\mathcal{I}$. Then this (restricted) differential system
\eqref{150417:0935} only has the {\it one} solution
\begin{equation}
\vy(x)=e^{-(x-x_0)\vA}\cdot \vy_0,\;\;\text{for all}\;\; x\in
\mathcal{I}.
\end{equation}
In particular, if $\vy_0=\boldsymbol{0}$ then
$\vy(x)=\boldsymbol{0}$ is the only solution for all  $x\in
\mathcal{I}$.

\vspace{1em}\noindent {\it Proof:} Let $\vy=\vy(x)$ be {\it any}
solution which satisfies the initial value problem
\eqref{150417:0935} in the given interval $\mathcal{I}$. Then we
can formulate the obvious relation
\begin{equation*}
\frac{d}{dx}\left(e^{(x-x_0)\vA}\cdot\vy\right)= \left(\vA\cdot
e^{(x-x_0)\vA}\right)\cdot\vy + e^{(x-x_0)\vA}\cdot\vy^\prime =
\left(\vA\cdot e^{(x-x_0)\vA}\right)\cdot\vy -
e^{(x-x_0)\vA}\cdot\left(\vA\cdot\vy\right) =\boldsymbol{0},
\end{equation*}
since the infinite matrix $\vA$ commutes with itself, i.e.
$[\vA,\vA^n]=\boldsymbol{0}$ for all $n\in\mathbb{N}$. This
relation then implies that
\begin{equation}
e^{(x-x_0)\vA}\cdot\vy(x)=\vc,\;\;\text{for
all}\;\;x\in\mathcal{I},
\end{equation}
where $\vc$ is some integration constant. But since $\vy$
satisfies the initial condition $\vy(x_0)=\vy_0$, we obtain the
result that $\vc=\vy_0$ and that thus the considered solution
$\vy$ can only have the unique~form:
\begin{equation}
\hspace{4.6cm}\vy(x)=e^{-(x-x_0)\vA}\cdot \vy_0,\;\;\text{for
all}\;\; x\in \mathcal{I}.\hspace{3.25cm}\square
\end{equation}

\pagebreak[4]

\begin{figure}[h!]
\begin{subfigure}[c]{.48\linewidth}
\centering
\includegraphics[width=.91\textwidth]{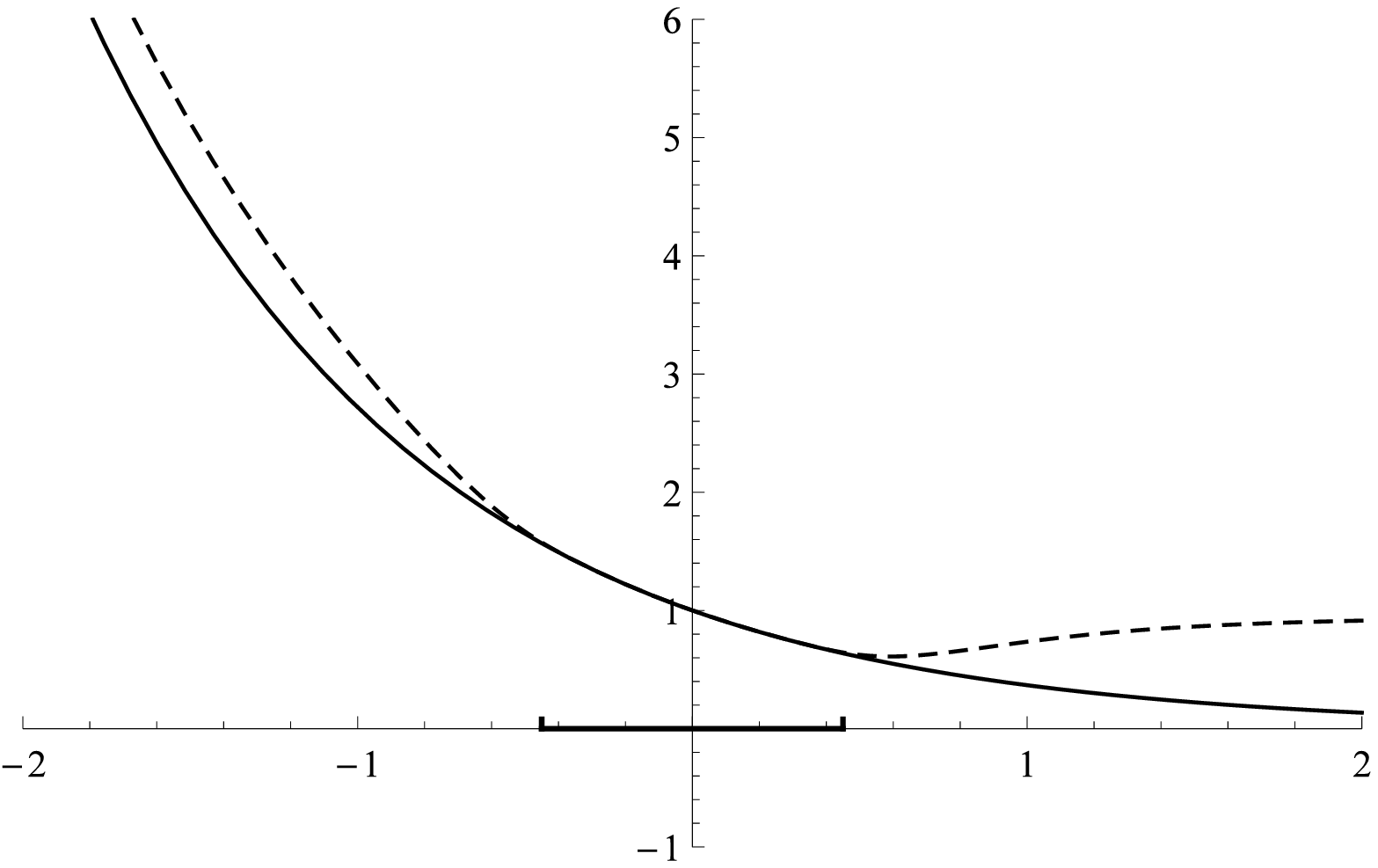}
\caption{$n=1,\;\: \gamma=1,\;\: |\mathcal{I}_1|\sim 0.9$}
\label{F1a}
\end{subfigure}
\begin{subfigure}[c]{.48\linewidth}
\centering
\includegraphics[width=.91\textwidth]{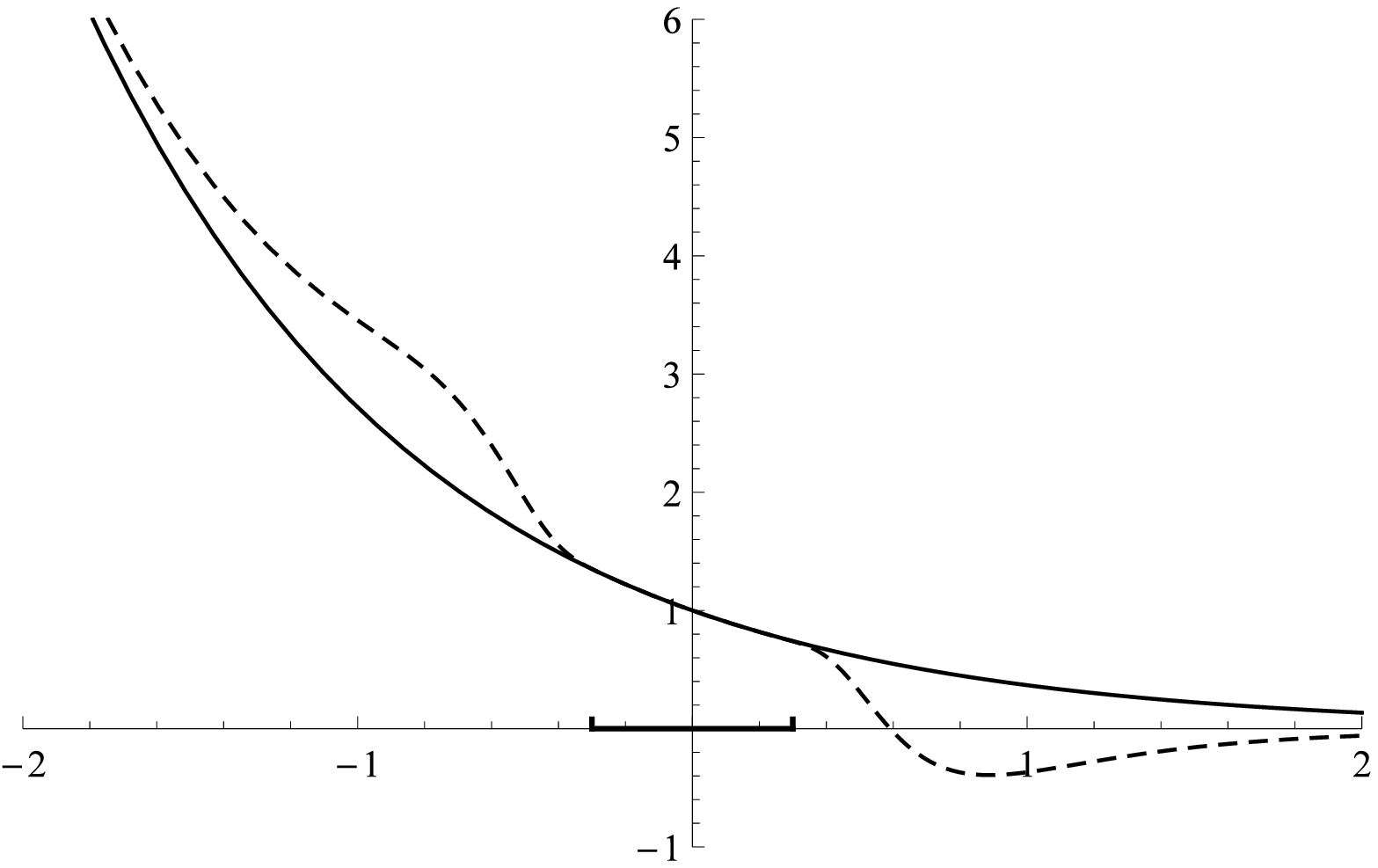}
\caption{$n=2,\;\: \gamma=1,\;\: |\mathcal{I}_2|\sim 0.6$}
\label{F1b}
\end{subfigure}
\begin{subfigure}[c]{.48\linewidth}
\centering
\includegraphics[width=.91\textwidth]{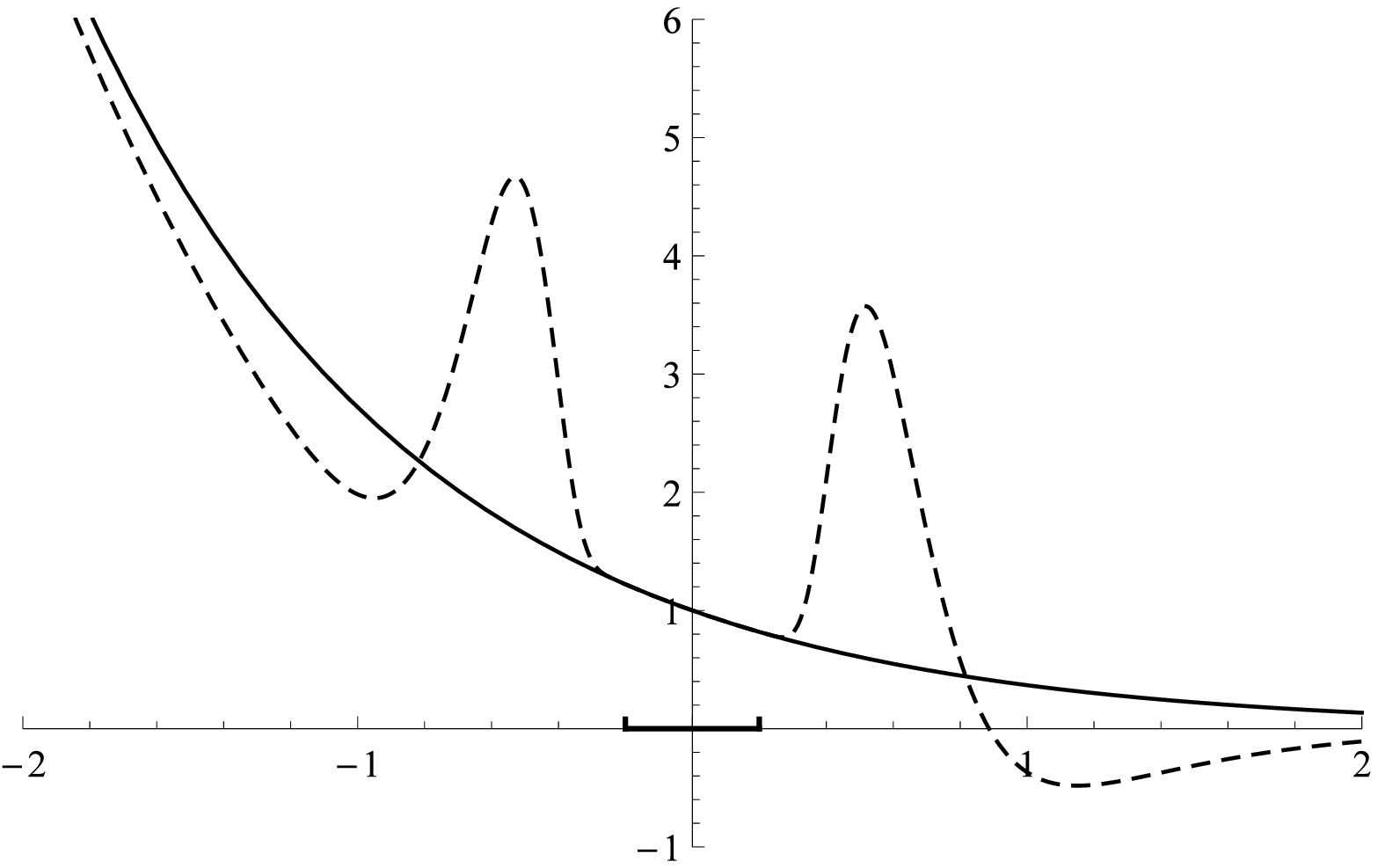}
\caption{$n=3,\;\: \gamma=1,\;\: |\mathcal{I}_3|\sim 0.4$}
\label{F1c}
\end{subfigure}
\begin{subfigure}[c]{.48\linewidth}
\centering
\includegraphics[width=.91\textwidth]{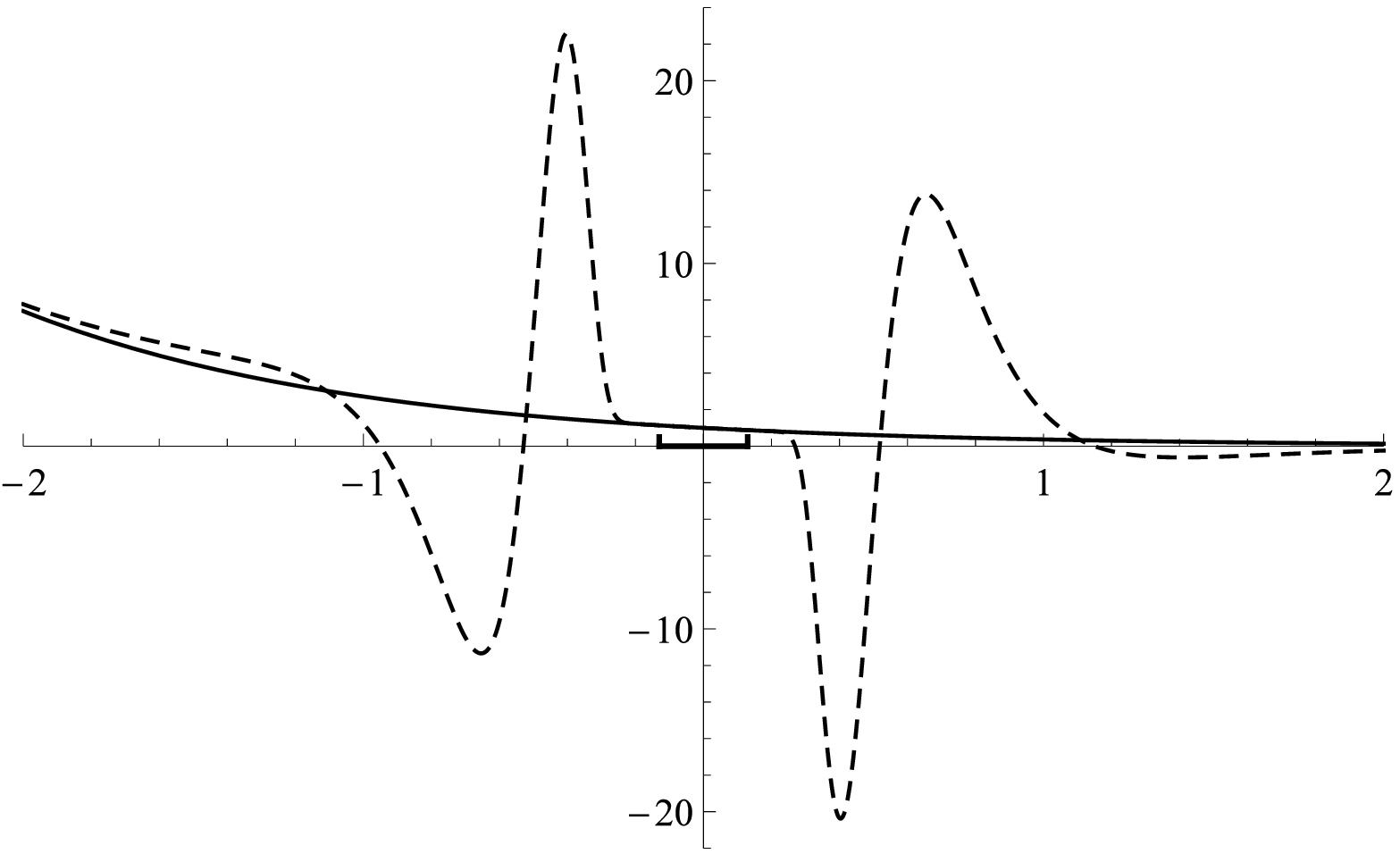}
\caption{$n=4,\;\: \gamma=1,\;\: |\mathcal{I}_4|\sim 0.3$}
\label{F1d}
\end{subfigure}
\caption{Plots of the first four solutions of the initial value
problem \eqref{150417:1123}. The solid lines display the solutions
$y_n^\text{I}$ \eqref{150418:1428} and the dashed lines the
solutions $y_n^\text{II}$ \eqref{150418:1429} for $\gamma=1$. For
each order $n$, the highlighted region on the $x$-axis indicates
the local uniqueness interval $\mathcal{I}_n$ where
$y_n^\text{I}=y_n^\text{II}$. The size of each interval
$|\mathcal{I}_n|$ decreases as the order $n$ of the solution
increases. In the limit $n\rightarrow\infty$ the corresponding
interval narrows down to point-size, i.e.
$\lim_{n\to\infty}|\mathcal{I}_n|\rightarrow 0$. Hence, the size
of the {\it common} uniqueness interval $\mathcal{I}$ of the
initial value problem \eqref{150417:1123}, which is the
intersection of all intervals $\mathcal{I}=\bigcap_{n=1}^\infty
\mathcal{I}_n$, thus converges to point-size too.} \label{F1}
\end{figure}
\begin{figure}[h!]
\begin{subfigure}[c]{.48\linewidth}
\centering
\includegraphics[width=.91\textwidth]{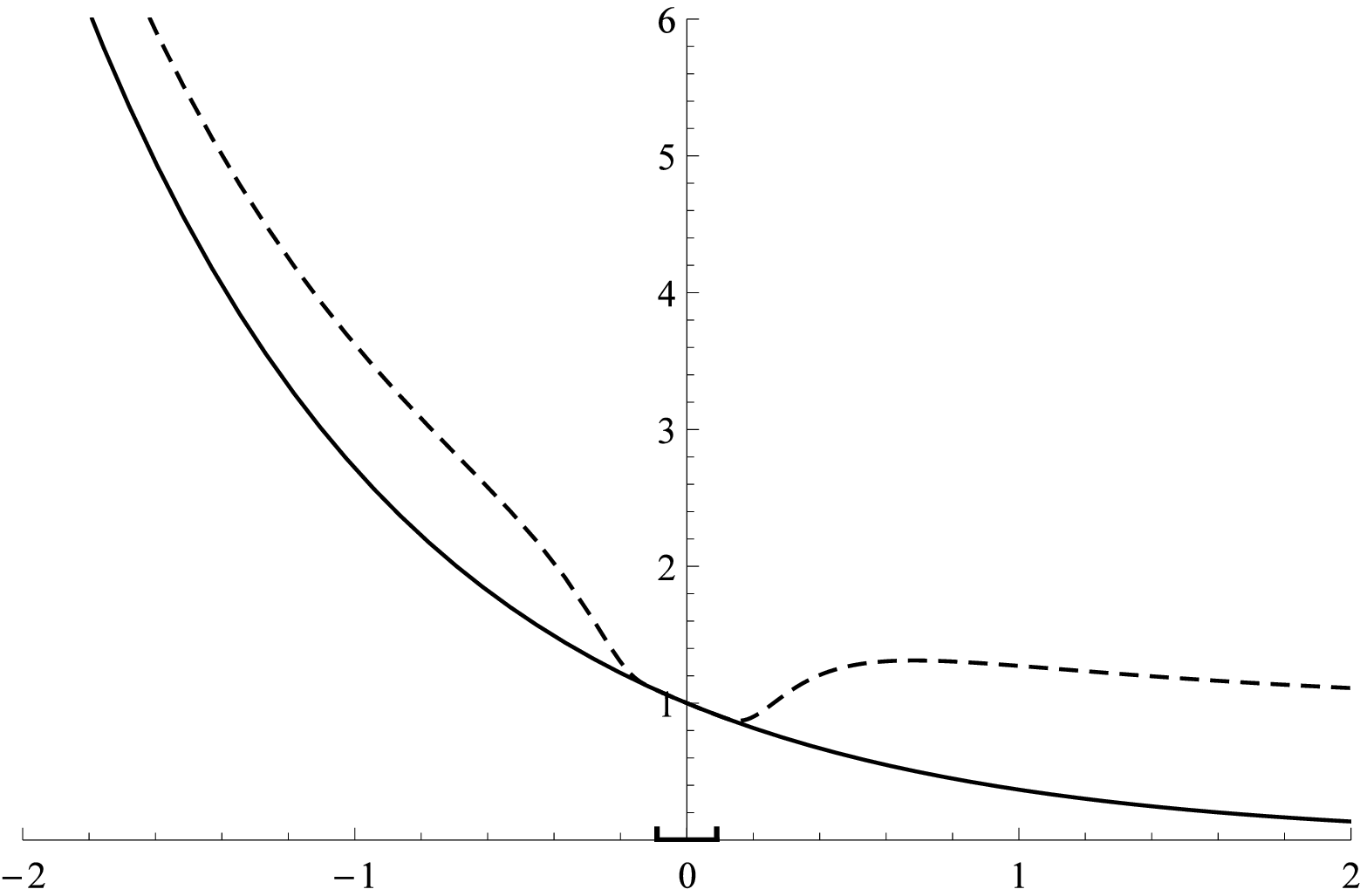}
\caption{$n=1,\;\: \gamma=0.1,\;\: |\mathcal{I}_1|\sim 0.2$}
\label{F2a}
\end{subfigure}
\begin{subfigure}[c]{.48\linewidth}
\centering
\includegraphics[width=.91\textwidth]{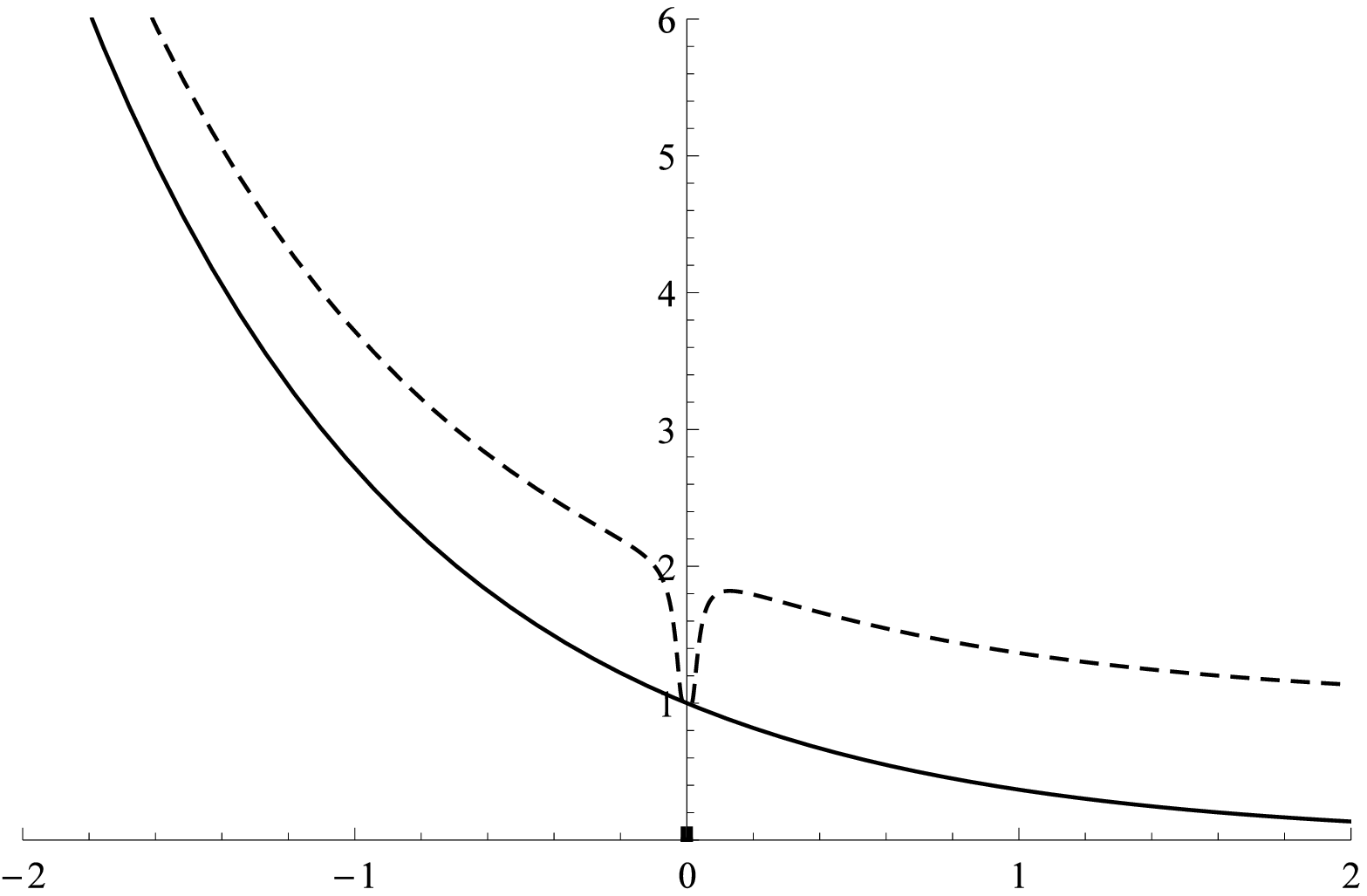}
\caption{$n=1, \;\: \gamma=0.001,\;\: |\mathcal{I}_1|\sim 0.02$}
\label{F2b}
\end{subfigure}
\caption{Plots of the first order solutions $y_1^\text{I}$
(\eqref{150418:1428}, solid lines) and $y_1^\text{II}$
(\eqref{150418:1429}, dashed lines) for decreasing $\gamma$.
Hence, for $\gamma\rightarrow 0$ the size of the local uniqueness
interval $\mathcal{I}_{n^*}$ for each arbitrary but fixed order
$n=n^*$ diminishes to point-size.} \label{F2}
\end{figure}

\pagebreak[4]\noindent {\it Example:} In how narrow this local
uniqueness interval $\mathcal{I}$ can be made up to point-size, we
want to demonstrate at the following specific initial value
problem \eqref{150417:0935}
\begin{equation}
\vy^\prime=-\vA\cdot\vy,\;\;\text{with}\;\; \vy(0)=\boldsymbol{1},
\label{150417:1123}
\end{equation}
in some {\it local} interval $\mathcal{I}\subset \mathbb{R}$ for
$x$ around the initial point $x_0=0$. Of course, {\it globally},
i.e. for all $x\in\mathbb{R}$, the solution of the initial value
problem \eqref{150417:1123} is not necessarily unique. Indeed, at
least two global solutions $\vy=(y_n)_{n\in\mathbb{N}}$ can be
found, e.g.
\begin{align}
y_n^\text{I} & =e^{-x}, \;\;\text{for all}\;\; n\geq 1,\label{150418:1428}\\[0.75em]
y_{n}^{\text{II}} & = \begin{cases}\,
e^{-x}+e^{-\frac{\gamma}{x^2}}, \;\;\text{for}\;\; n= 1,\; \gamma
>0,\\[0.5em]
\, (-1)^{n-1}{\displaystyle\frac{d^{n-1}
y_1^{\text{II}}}{dx^{n-1}}}, \;\;\text{for}\;\; n\geq
2,\end{cases} \label{150418:1429}
\end{align}

\noindent which both satisfy \eqref{150417:1123} for all
$n\in\mathbb{N}$, and all $x\in\mathbb{R}$. Figure \ref{F1} and
\ref{F2} shows this for constant and different $\gamma$
respectively.

\section{Remapping of the general solutions' integration constant\label{D}}

\subsection{Reparametrization of solution $n=1$\label{D1}}

\vspace{-1.5em}
\begin{align}
\tilde{y}_1 & = e^{\varepsilon}y_1\nonumber\\
& =e^{\varepsilon}\sum_{k=0}^\infty c_{1+k}\frac{(-1)^k}{k!}\, x^k
= e^{\varepsilon}\sum_{k=0}^\infty
c_{1+k}\frac{(-1)^k}{k!}\left(\frac{\tilde{x}}{1+\varepsilon\tilde{x}}\right)^k\nonumber\\
& =e^{\varepsilon}\sum_{k=0}^\infty
c_{1+k}\frac{(-1)^k}{k!}\left(\,\sum_{l=0}^\infty
(-1)^l\frac{k\,(k+l)!}{k!\,(k+l)}\frac{\varepsilon^l}{l!}
\frac{\tilde{x}^k(\tilde{x}-a)^{l}}{(1+a\,\varepsilon)^{k+l}}\right),\;\;
\text{for}\;\; \bigg\vert\,
\frac{(\tilde{x}-a)\,\varepsilon}{1+a\,\varepsilon}\,\bigg\vert <
1 ,\;\;
a\neq -\frac{1}{\varepsilon},\nonumber\\
& = e^{\varepsilon}\sum_{k=0}^\infty
c_{1+k}\frac{(-1)^k}{k!}\left(\,\sum_{l=0}^\infty\,\sum_{q=0}^l
\frac{(-1)^{l+q}\, a^q}{(1+a\,\varepsilon)^{k+l}}
\binom{l}{q}\frac{k\,(k+l)!}{k!\,(k+l)}\frac{\varepsilon^l}{l!} \,
\tilde{x}^{k+l-q}\right)
\nonumber\\
& = e^{\varepsilon}\sum_{k=0}^\infty
c_{1+k}\frac{(-1)^k}{k!}\left(\,\sum_{m=0}^\infty\,\sum_{r=0}^\infty
\frac{(-1)^{m}\, a^r}{(1+a\,\varepsilon)^{k+m+r}}
\binom{m+r}{r}\frac{k\,(k+m+r)!}{k!\,(k+m+r)}\frac{\varepsilon^{m+r}}{(m+r)!}
\, \tilde{x}^{k+m}\right)
\nonumber\\
& = \sum_{k=0}^\infty\, \sum_{m=0}^\infty\, \sum_{r=0}^\infty
e^{\varepsilon}c_{1+k}\frac{(-1)^{k+m}\,
a^r\,\varepsilon^{m+r}}{(1+a\,\varepsilon)^{k+m+r} k!\,
(m+r)!}\binom{m+r}{r}\frac{k\,(k+m+r)!}{k!\,(k+m+r)}\,\tilde{x}^{k+m}\nonumber\\
& =  \sum_{i=0}^\infty \, \sum_{j=0}^i \, \sum_{r=0}^\infty
e^{\varepsilon} c_{1+j} \frac{(-1)^{i}\,
a^r\,\varepsilon^{i-j+r}}{(1+a\,\varepsilon)^{i+r} j!\,
(i-j+r)!}\binom{i-j+r}{r}\frac{j\, (i+r)!}{j!\,
(i+r)}\,\tilde{x}^{i}\nonumber\\
& = \sum_{i=0}^\infty \left(\,\sum_{j=0}^i \, \sum_{r=0}^{\infty}
e^{\varepsilon} c_{1+j} \frac{i!\,
a^r\,\varepsilon^{i-j+r}}{(1+a\,\varepsilon)^{i+r} j!\, r!\,
(i-j)!}\frac{j\, (i+r)!}{j!\,
(i+r)}\right)\frac{(-1)^i}{i!}\,\tilde{x}^i\nonumber\\
& = \sum_{i=0}^\infty \left(\,\sum_{j=0}^i e^{\varepsilon} c_{1+j}
\frac{i!\, \varepsilon^{i-j}}{j!\, (i-j)!}\frac{i!\, j}{i\, j!}
\sum_{r=0}^\infty
\frac{(a\,\varepsilon)^r}{(1+a\,\varepsilon)^{i+r}\, r!}
\frac{i\, (i+r)!}{i!\, (i+r)}\right)\frac{(-1)^i}{i!}\,\tilde{x}^i\nonumber\\
& = \sum_{i=0}^\infty \left(\,\sum_{j=0}^i e^{\varepsilon} c_{1+j}
\frac{i!\,\varepsilon^{i-j}}{j!\, (i-j)!}\frac{i!\, j}{i\,
j!}\cdot 1\right)\frac{(-1)^i}{i!}\,\tilde{x}^i,\;\;
\text{for}\;\; \bigg\vert\,
\frac{a\,\varepsilon}{1+a\,\varepsilon}\,\bigg\vert <
1 ,\nonumber\\
& =: \sum_{i=0}^\infty
\tilde{c}_{1+i}\frac{(-1)^i}{i!}\,\tilde{x}^i,\;\; \text{for}\;\;
\vert\, \tilde{x}\varepsilon\,\vert < 1 , \label{150409:1333}
\end{align}
where we made use of the Cauchy product rule in both directions:
\begin{equation}
\sum_{k=0}^\infty\, \sum_{l=0}^\infty f_k\cdot g_l \cdot h_{k+l} =
\sum_{i=0}^\infty \, \sum_{j=0}^i f_j\cdot g_{i-j}\cdot h_i.
\end{equation}
Note that the reparametrization of the integration constant
\begin{equation}
c_{1+i}\mapsto \tilde{c}_{1+i}=
\begin{cases}
\; e^\varepsilon c_1,\;\;\text{for}\;\; i=0,\\[0.4em]
\; {\displaystyle\sum_{j=0}^i e^{\varepsilon} c_{1+j}
\frac{i!\,\varepsilon^{i-j}}{j!\, (i-j)!}\frac{i!\, j}{i\, j!}},
\;\;\text{for}\;\; i\geq 1,
\end{cases}
\label{150409:1325}
\end{equation}
is independent of the expansion point $a$ for all $i\geq 0$. In
particular, relation \eqref{150409:1325} represents the
reparametrization for $a=0$, which explains the third and last
constraint $\vert\, \tilde{x}\varepsilon\,\vert < 1$ in
\eqref{150409:1333}. Note that in vector form relation
\eqref{150409:1325} can be condensed to
\begin{equation}
\vc\mapsto\tilde{\vc}=\vG(0,\varepsilon)\cdot \vc,
\label{150502:1359}
\end{equation}
where the infinite matrix $\vG$ is defined by \eqref{150407:1002}.

\subsection{Reparametrization of all remaining solutions $n\geq 2$\label{D2}}

\vspace{-2.5em}
\begin{align}
\tilde{y}_n & = \sum_{k=1}^{n-1}
B_{n,k}\,\varepsilon^{n-k-1}(1-\varepsilon x)^{n+k-1}\,
e^{\varepsilon}\, y_{k+1},\;\;\text{for all}\;\; n\geq 2,\nonumber\\
& = \sum_{k=1}^{n-1} B_{n,k}\,\varepsilon^{n-k-1}(1-\varepsilon
x)^{n+k-1}\, e^{\varepsilon}\,
\sum_{l=0}^\infty c_{k+1+l}\frac{(-1)^l}{l!}x^l\nonumber\\
& = \sum_{k=1}^{n-1} B_{n,k}\,\varepsilon^{n-k-1}
\left(1-\frac{\varepsilon
\tilde{x}}{1+\varepsilon\tilde{x}}\right)^{n+k-1}
e^{\varepsilon}\, \sum_{l=0}^\infty
c_{k+1+l}\frac{(-1)^l}{l!}\left(\frac{\tilde{x}}{1+\varepsilon\tilde{x}}\right)^l\nonumber\\
& = \sum_{k=1}^{n-1}\, \sum_{l=0}^\infty
B_{n,k}\,\varepsilon^{n-k-1} e^{\varepsilon} c_{k+1+l}
\frac{(-1)^l}{l!}
\frac{\tilde{x}^l}{(1+\varepsilon\tilde{x})^{n+k+l-1}}
\nonumber\\
& =\sum_{k=1}^{n-1}\, \sum_{l=0}^\infty
B_{n,k}\,\varepsilon^{n-k-1} e^{\varepsilon} c_{k+1+l}
\frac{(-1)^l}{l!}\nonumber\\
& \hspace{0.5cm}\cdot \left(\,\sum_{m=0}^\infty
(-1)^m\frac{(n+k+l+m-2)!}{(n+k+l-2)!}\frac{\varepsilon^m}{m!}
\frac{\tilde{x}^{l}(\tilde{x}-a)^m}{(1+a\,\varepsilon)^{n+k+l+m-1}}\right),\nonumber\\
& \hspace{1.275cm}\text{for}\;\; \bigg\vert\,
\frac{(\tilde{x}-a)\,\varepsilon}{1+a\,\varepsilon}\,\bigg\vert <
1 ,\;\; a\neq -\frac{1}{\varepsilon},\nonumber\\
& =\sum_{k=1}^{n-1}\, \sum_{l=0}^\infty
B_{n,k}\,\varepsilon^{n-k-1} e^{\varepsilon} c_{k+1+l}
\frac{(-1)^l}{l!}\nonumber\\
& \hspace{0.5cm}\cdot \left(\,\sum_{m=0}^\infty\, \sum_{q=0}^m
\frac{(-1)^{m+q}\,
a^q}{(1+a\,\varepsilon)^{n+k+l+m-1}}\binom{m}{q}\frac{(n+k+l+m-2)!}{(n+k+l-2)!}
\frac{\varepsilon^m}{m!}\,\tilde{x}^{l+m-q}\right)\qquad\qquad\quad\nonumber
\end{align}
\pagebreak[4]
\begin{align}
\tilde{y}_n & =\sum_{k=1}^{n-1}\, \sum_{l=0}^\infty
B_{n,k}\,\varepsilon^{n-k-1} e^{\varepsilon} c_{k+1+l}
\frac{(-1)^l}{l!}\nonumber\\
& \hspace{0.5cm}\cdot \left(\,\sum_{p=0}^\infty\,
\sum_{r=0}^\infty \frac{(-1)^{p}\,
a^r}{(1+a\,\varepsilon)^{n+k+l+p+r-1}}\binom{p+r}{r}\frac{(n+k+l+p+r-2)!}{(n+k+l-2)!}
\frac{\varepsilon^{p+r}}{(p+r)!}\,\tilde{x}^{l+p}\right)\nonumber\\
& =\sum_{i=0}^\infty\,\sum_{j=0}^i\,\sum_{k=1}^{n-1}
B_{n,k}\,\varepsilon^{n-k-1} e^{\varepsilon} c_{k+1+j}
\frac{(-1)^j}{j!}\nonumber\\
& \hspace{0.5cm}\cdot \left(\, \sum_{r=0}^\infty
\frac{(-1)^{i-j}\,
a^r}{(1+a\,\varepsilon)^{n+k+i+r-1}}\binom{i-j+r}{r}\frac{(n+k+i+r-2)!}{(n+k+j-2)!}
\frac{\varepsilon^{i-j+r}}{(i-j+r)!}\,\tilde{x}^{i}\right)\nonumber\\
& =\sum_{i=0}^\infty\left(\,\sum_{k=1}^{n-1}\,\sum_{j=0}^i
B_{n,k}\,\varepsilon^{n+i-j-k-1} e^{\varepsilon} c_{k+1+j}
\frac{i!}{j!\,(i-j)!}\frac{(n+k+i-2)!}{(n+k+j-2)!}\right)\frac{(-1)^i}{i!}\,\tilde{x}^i\nonumber\\
& \hspace{0.5cm}\cdot \left(\, \sum_{r=0}^\infty \frac{
(a\,\varepsilon)^r}{(1+a\,\varepsilon)^{n+k+i+r-1}\,
r!}\frac{(n+k+i+r-2)!}{(n+k+i-2)!} \right),\;\; \text{and if}\;\;
\bigg\vert\, \frac{a\,\varepsilon}{1+a\,\varepsilon}\,\bigg\vert <
1, \;\;\text{then:}\nonumber\\
& =\sum_{i=0}^\infty\left(\,\sum_{k=1}^{n-1}\,\sum_{j=0}^i
B_{n,k}\,\varepsilon^{n+i-j-k-1} e^{\varepsilon} c_{k+1+j}
\frac{i!}{j!\,(i-j)!}\frac{(n+k+i-2)!}{(n+k+j-2)!}\right)\frac{(-1)^i}{i!}\,\tilde{x}^i\nonumber\\
& =: \sum_{i=0}^\infty
\tilde{c}_{n+i}\frac{(-1)^i}{i!}\,\tilde{x}^i, \;\; \text{for}\;\;
\vert\, \tilde{x}\varepsilon\,\vert < 1,\;\; \text{and}\;\; n\geq
2.\label{150409:2143}
\end{align}

\noindent Note that the reparametrization of the integration
constant
\begin{equation}
c_{n+i}\mapsto \tilde{c}_{n+i}= \sum_{k=1}^{n-1}\,\sum_{j=0}^i
B_{n,k}\,\varepsilon^{n+i-j-k-1} e^{\varepsilon} c_{k+1+j}
\frac{i!}{j!\,(i-j)!}\frac{(n+k+i-2)!}{(n+k+j-2)!},\;\; n\geq 2,
\label{150409:2319}
\end{equation}
is again independent of the expansion point $a$ for all $i\geq 0$
and $n\geq 2$, and that it basically represents the exact
reparametrization for $a=0$, which thus again explains the third
and last constraint $\vert\, \tilde{x}\varepsilon\,\vert < 1$ in
\eqref{150409:2143}. Obviously, when evaluated,
\eqref{150409:2319} must give the same result for the transformed
integration constant
$\tilde{\vc}^T=(\tilde{c}_1,\tilde{c}_2,\tilde{c}_3,\dots
\tilde{c}_n,\dots)$ as \eqref{150409:1325}, which is
\eqref{150502:1359}.

\section{Inverse infinite group matrix\label{E}}

This section demonstrates how the inverse of the infinite group
matrix $\vG$ \eqref{150407:1002} is constructed. Since, by
construction, the group transformation $\mathsf{L}_2$
\eqref{150407:1002} is based on an additive composition law of the
group parameter $\varepsilon$, the inverse transformation of
$\mathsf{L}_2$ \eqref{150407:1002} is thus given as
\begin{equation}
\mathsf{L}_2^{-1}: \;\;\; x=\frac{\tilde{x}}{1+\varepsilon
\tilde{x}},\;\;\;\vy=\vG(\tilde{x},-\varepsilon)\cdot\tilde{\vy}.
\label{150413:1359}
\end{equation}
And since the transformation $\vy\mapsto\tilde{\vy}$ in
$\mathsf{L}_2$ \eqref{150407:1002} can be formally written as
\begin{equation}
\vy=\vG^{-1}(x,\varepsilon)\cdot\tilde{\vy},
\end{equation}
the infinite inverse matrix $\vG^{-1}$ is thus defined as
\begin{equation}
\vG^{-1}(x,\varepsilon)=\vG\big({\textstyle\frac{x}{1-\varepsilon
x}},-\varepsilon\big). \label{150519:0947}
\end{equation}

\section{Explicit forms and graphs of the solutions $\vy^A$ and $\vy^B$\label{F}}

\begin{figure}
\begin{subfigure}[c]{.48\linewidth}
\centering
\includegraphics[width=.91\textwidth]{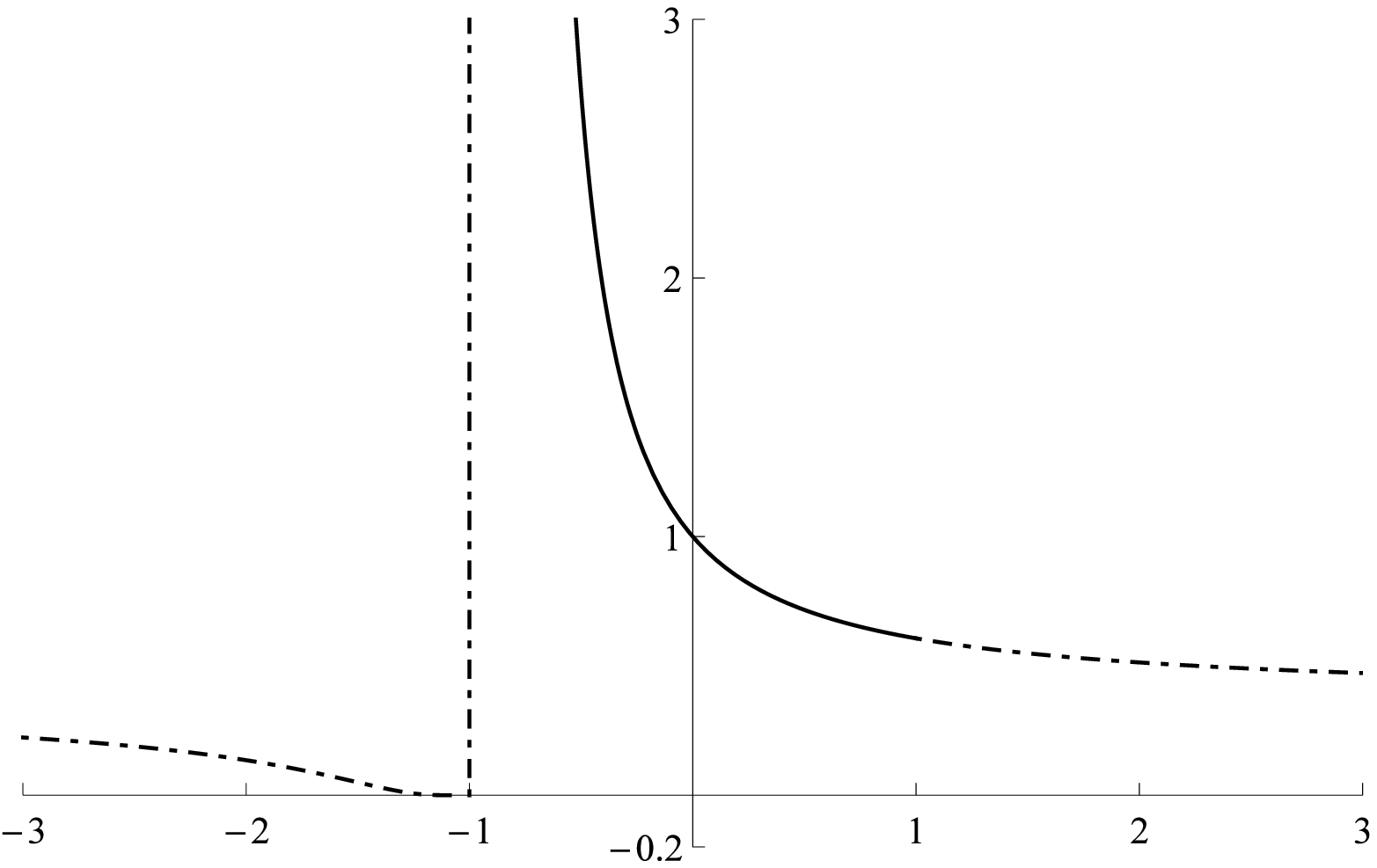}
\caption{$n=1,\;\: y_1^A\:(\text{solid-line})\subset y_1^B$}
\label{F3a}
\end{subfigure}
\begin{subfigure}[c]{.48\linewidth}
\centering
\includegraphics[width=.91\textwidth]{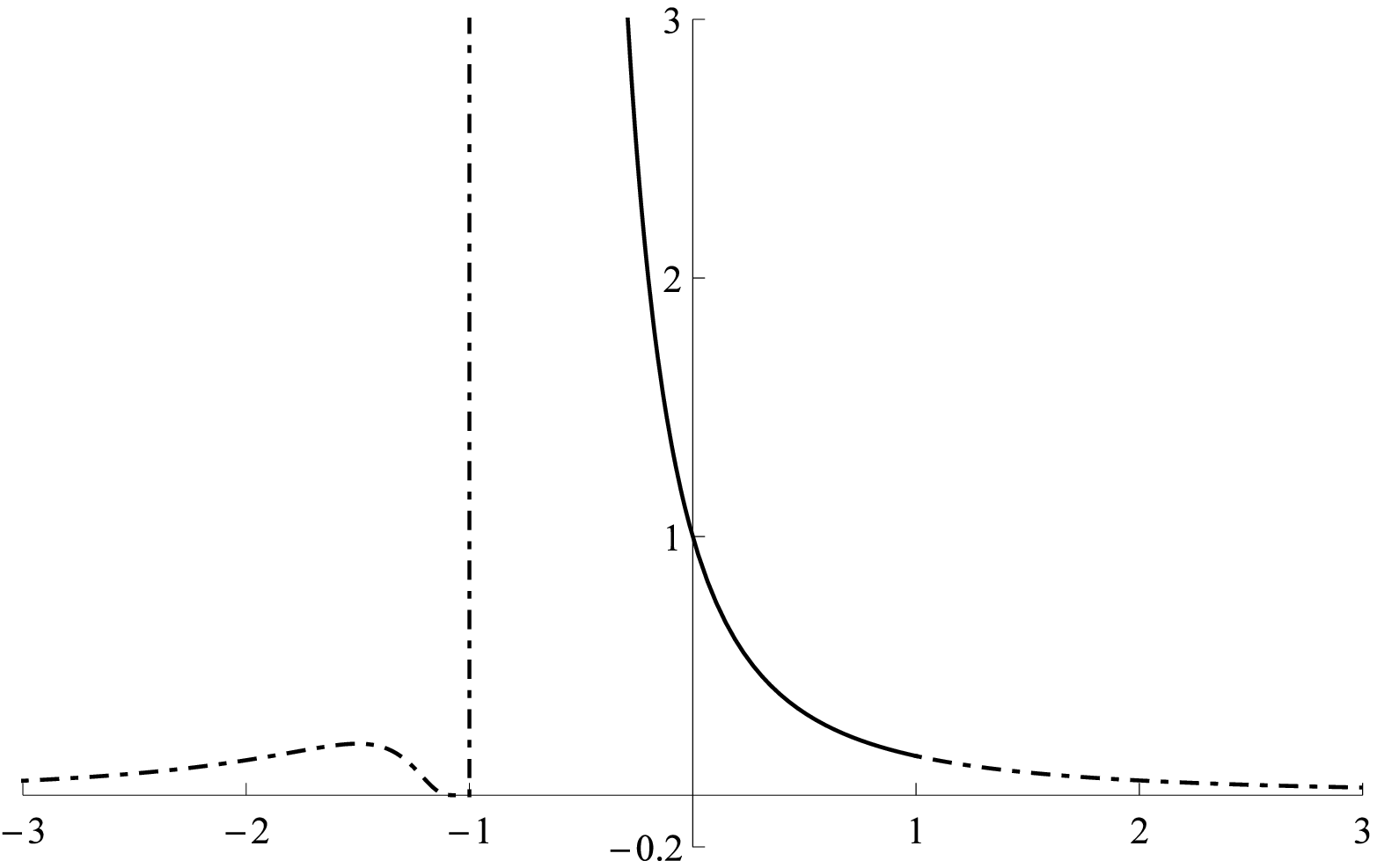}
\caption{$n=2,\;\: y_2^A\:(\text{solid-line})\subset y_2^B$}
\label{F3b}
\end{subfigure}
\begin{subfigure}[c]{.48\linewidth}
\centering
\includegraphics[width=.91\textwidth]{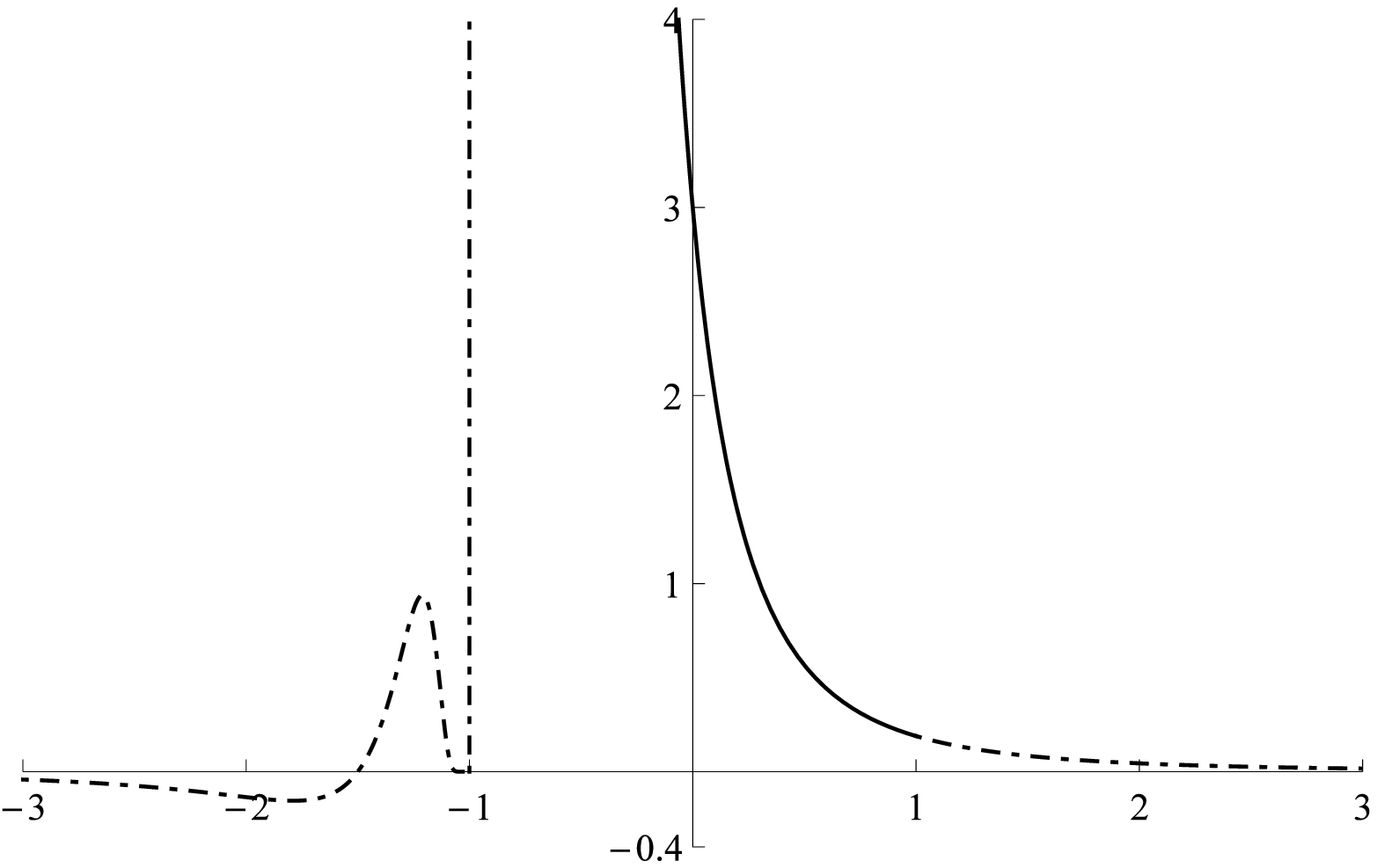}
\caption{$n=3,\;\: y_3^A\:(\text{solid-line})\subset y_3^B$}
\label{F3c}
\end{subfigure}
\begin{subfigure}[c]{.48\linewidth}
\centering
\includegraphics[width=.91\textwidth]{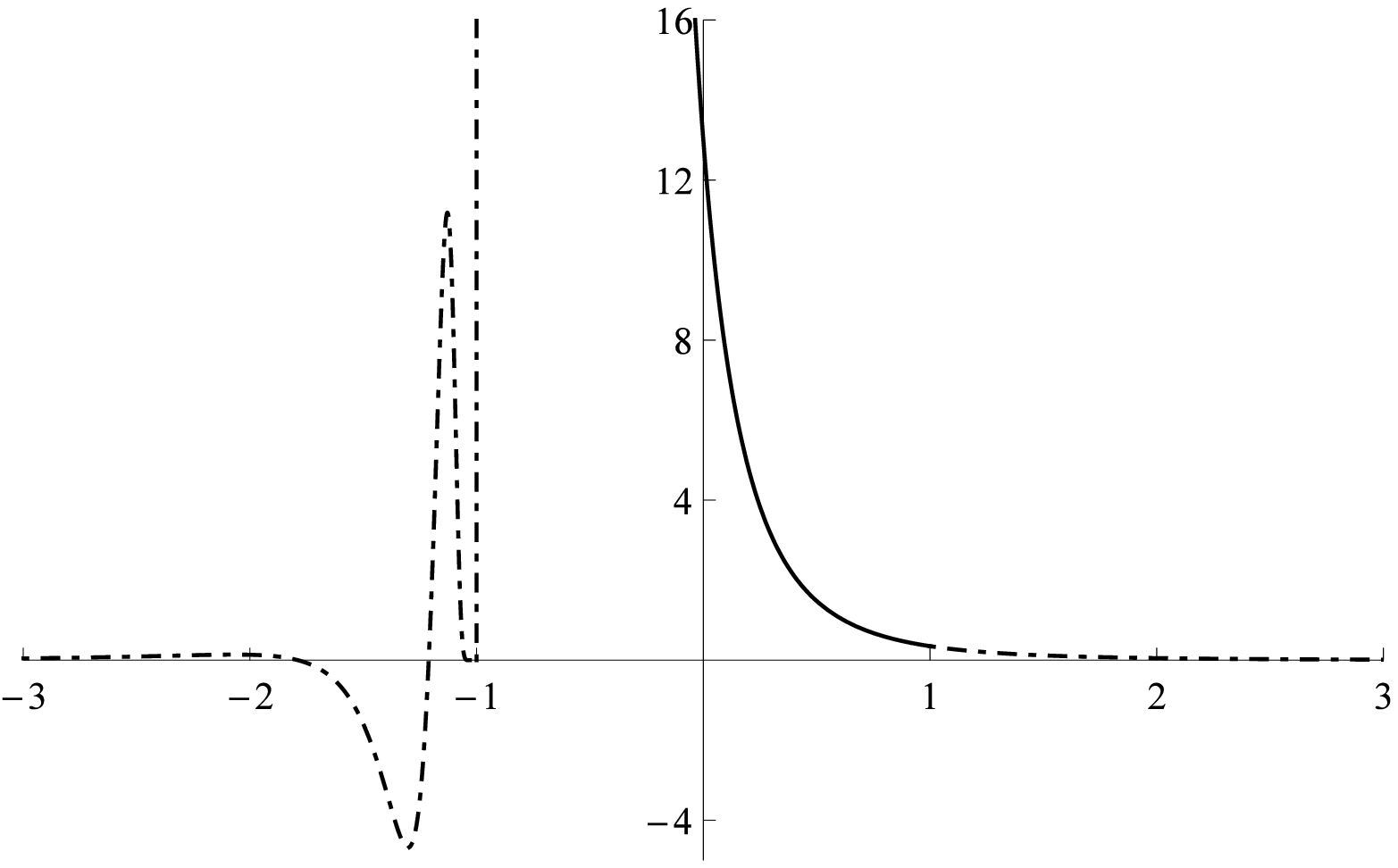}
\caption{$n=4,\;\: y_4^A\:(\text{solid-line})\subset y_4^B$}
\label{F3d}
\end{subfigure}
\caption{Plots of the first four solutions of the initial value
problem \eqref{150519:1034}. The solid lines display the solutions
$y_n^A$ \eqref{150413:1430}, while the solutions $y_n^B$
\eqref{150413:1431} are given by the solid lines along with the
extensions displayed by the dashed lines. The initial condition
was set $\vy_0=\vG(0,\varepsilon)\cdot \vc$, with
$\vc=e^{-\varepsilon}\cdot\boldsymbol{1}$, and $\varepsilon=1$.\\[-0.92em]}
\label{F3}
\end{figure}

The explicit componential form of the primary solution $\vy^A$
\eqref{150413:1430} is given by \eqref{150401:1925}
\begin{equation}
y^A_n(x)=\sum_{k=0}^\infty y_{(0)n+k}\frac{(-1)^k}{k!}\, x^k,\;\;
n\geq 1,
\end{equation}
while to bring the transformed solution $\vy^B$
\eqref{150413:1431} into its corresponding componential form, one
first has to recognize that its matrix-vector structure is
iteratively composed as
\begin{equation}
\vy^B(x)=\vG(x^* ,\varepsilon)\cdot\vy^*(x^*),\;\;\text{with}\;\;
x^*=\frac{x}{1+\varepsilon x}, \label{150413:1547}
\end{equation}
where
\begin{equation}
\vy^*(x^*)= e^{-x^*\vA}\cdot\vy_0^*, \label{150413:1548}
\end{equation}
and
\begin{equation}
\vy_0^*=\vG(0,-\varepsilon)\cdot \vy_0. \label{150413:1549}
\end{equation}
Then, according to \eqref{150403:1804}, the componential form of
\eqref{150413:1547} is given as
\begin{equation}
y_n^B(x)=
\begin{cases}
\; e^{\varepsilon}y_1^*(x^*),\;\;\text{for}\;\; n=1,  \\[0.5em]
\; {\displaystyle\sum_{k=1}^{n-1}}
B_{n,k}\,\varepsilon^{n-k-1}(1-\varepsilon x^*)^{n+k-1}\,
e^{\varepsilon}\, y^*_{k+1}(x^*),\;\;\text{for}\;\; n\geq 2,
\end{cases}
\label{150413:1647}
\end{equation}

\noindent where \eqref{150413:1548}, according to
\eqref{150401:1925}, has the form
\begin{equation}
y^*_q(x^*)=\sum_{l=0}^\infty y^{*}_{(0)q+l}\frac{(-1)^l}{l!}\,
(x^*)^l,\;\; q\geq 1, \label{150413:1646}
\end{equation}
and \eqref{150413:1549}, again according to \eqref{150403:1804},
but now for $\vy=\vy_0$, goes over into
\begin{equation}
y_{(0)r}^*=
\begin{cases}
\; e^{-\varepsilon}y_{(0)1},\;\;\text{for}\;\; r=1,  \\[0.5em]
\; {\displaystyle\sum_{m=1}^{r-1}}
B_{r,m}\,(-\varepsilon)^{r-m-1}\, e^{-\varepsilon}\,
y_{(0)m+1},\;\;\text{for}\;\; r\geq 2,
\end{cases}
\end{equation}

\noindent which then needs to be inserted back into
\eqref{150413:1646}, and this result again back into
\eqref{150413:1647} to finally give the componential form of
$\vy^B$ \eqref{150413:1431}. For a fixed set of initial
conditions, Figure~\ref{F3} displays the solutions $\vy^A$ and
$\vy^B$. The convergence domain for each solution for different
initial conditions is given in Table \ref{T1}.

\vspace{3em}\pagebreak[4]

\begin{table}[h!]
\begin{center}
\begin{tabular}{| c | c | c |}\hline & & \\[-0.50em]
$\vy_0=(y_{(0)i})_{i\in\mathbb{N}}$ & $\;\vy^A\;$ & $\;\vy^B\;$\\[0.5em]
\hline & & \\[-0.5em] $y_{(0)i}={\displaystyle\sum_{j=0}^\infty} G_{ij}(0,\varepsilon)\, c_j$,
with $c_j=\begin{cases}\, \alpha_1 j^n,\;\text{for any finite}\;
n\in\mathbb{R}\negthickspace\negthickspace\negthickspace\\[0.5em] \, \alpha_2$, for all
$j\in\mathbb{N}\\[0.5em] \, {\displaystyle\frac{\alpha_3}{i!}}
\end{cases}$ & $\vert\, x\,\varepsilon\,\vert <1$
&
$x\in\mathbb{R}\backslash\{{\textstyle -\frac{1}{\varepsilon}}\}$\\[3.5em]
$y_{(0)i}=\alpha_4\, i!$ & $\vert\, x\,\vert <1$ & $\,\bigg\vert\,
{\displaystyle\frac{x(1-\varepsilon)}{1+\varepsilon\, x}}
\,\bigg\vert <1$\\[1.5em]
$y_{(0)i}=\begin{cases}\, \alpha_5\, i^n,\;\text{for any finite}\;
n\in\mathbb{R}\\[0.5em] \, \alpha_6,\;\text{for all}\; i\in\mathbb{N} \\[0.5em] \,
{\displaystyle\frac{\alpha_7}{i!}}\end{cases}$ & $x\in\mathbb{R}$
& $\bigg\vert\,
{\displaystyle\frac{x\,\varepsilon}{1+\varepsilon\,
x}}\,\bigg\vert <1$\\[3.5em]
$y_{(0)i}=\alpha_8 (i!)^2$ & x\,=\,0 & x\,=\,0 \\[1.0em]
\hline
\end{tabular}
\caption{Convergence domains for the solutions $\vy^A$
\eqref{150413:1430} and $\vy^B$ \eqref{150413:1431} for a
collection of various different initial values $\vy_0$, where all
$\alpha$'s are arbitrary global constants. The domains were
determined by using the Cauchy-Hadamard root test.} \label{T1}
\end{center}
\vspace{-0.4em}
\end{table}

\section{Derivation of a general solution for the nonlinear
system\label{G}}

Given is the infinitely of first order coupled system of
Riccati-ODEs \eqref{150420:0806}
\begin{equation}
y_n^\prime-\frac{y_n}{x}=\frac{y_{n+1}^2}{x^3},\quad n\geq
1,\label{150421:0917}
\end{equation}
which, if a power series solution around some arbitrary expansion
point $x=a\in\mathbb{R}$ is sought, first should be transformed
into an adequate form. This is achieved by transforming the
function values as $y_n=x^2\cdot z_n$ to give the equivalent
differential system to \eqref{150421:0917}:
\begin{equation}
(x-a)\cdot z_n^\prime +a\cdot z_n^\prime + z_n = z_{n+1}^2.
\label{150423:1255}
\end{equation}
Inserting then the general Ansatz solution
\begin{equation}
z_n(x)=\sum_{k=0}^\infty \lambda_{n,k}\, (x-a)^k,\;\; n\geq 1,
\end{equation}
will turn this system of equations \eqref{150423:1255} into
\begin{align}
0 \; = & \;\; (x-a)\sum_{k=0}^\infty k\cdot\lambda_{n,k}\,
(x-a)^{k-1} +a
\sum_{k=0}^\infty k\cdot\lambda_{n,k}\, (x-a)^{k-1}\nonumber\\
&\;\; +\, \sum_{k=0}^\infty \lambda_{n,k}\, (x-a)^k-
\left(\;\sum_{k=0}^\infty \lambda_{n+1,k}\, (x-a)^k\right)\cdot
\left(\;\sum_{k=0}^\infty \lambda_{n+1,k}\,
(x-a)^k\right)\nonumber\\
= &\;\; \sum_{k=0}^\infty k\cdot\lambda_{n,k}\, (x-a)^{k} +a
\sum_{k=1}^\infty k\cdot\lambda_{n,k}\, (x-a)^{k-1}\nonumber\\
&\;\; +\, \sum_{k=0}^\infty \lambda_{n,k}\, (x-a)^k-
\sum_{k=0}^\infty\sum_{l=0}^k \lambda_{n+1,l}\lambda_{n+1,k-l}\,
(x-a)^{k}\nonumber\\
= &\;\; \sum_{k=0}^\infty (x-a)^k\left[k\cdot \lambda_{n,k}+
a\cdot (k+1)\cdot \lambda_{n,k+1}+\lambda_{n,k}-\sum_{l=0}^k
\lambda_{n+1,l}\cdot \lambda_{n+1,k-l}\right],
\end{align}
which, termwise equated, gives the following recurrence relation
for the expansion coefficients
\begin{equation}
k\cdot \lambda_{n,k}+ a\cdot (k+1)\cdot
\lambda_{n,k+1}+\lambda_{n,k}=\sum_{l=0}^k \lambda_{n+1,k-l}\cdot
\lambda_{n+1,l}\, ,\;\text{for all}\;\; n\geq 1,\, k\geq 0.
\end{equation}
For every arbitrary but fixed order $n$, the above relation
represents a 1-dimensional recurrence relation of first order
relative to index $k$, which can be uniquely solved by imposing
for all $n\geq 1$  at $k=0$ an initial condition
$\lambda_{n,0}=c_n$, where $c_n$ is some arbitrary constant. Note
that for the singular case $a=0$ the solution for the expansion
coefficients $\lambda_{n,k}$ will be different to those for all
$a\neq 0$.

\section{Proof that the general solution can be partially matched
to a special solution\label{H}}

The proposition is that for $a\neq 0$ the general solution $y_n$
\eqref{150423:1555} can only be matched to the special solution
$y_n^{(3)}$ \eqref{150423:2018} in the domain $|x-a|<|a|$. This
can be straightforwardly seen when performing the following two
steps: Firstly, equating these two solutions relative to $x^2$
\begin{equation}
\frac{y_n^{(3)}(x)}{x^2}=\frac{y_n(x)}{x^2},\;\;\text{for all}\;\:
n\geq 1,\label{150424:1819}
\end{equation}
will give the matching relation
\begin{equation}
\frac{1}{5^{2^{-n}}}\left[\,\prod_{k=0}^{n-1}
\left(1+\frac{1}{2^{k-2}}\right)^{2^{k-n}}\,\right]\cdot
x^{\frac{1}{2^{n-2}}}=\sum_{k=0}^\infty \lambda_{n,k}\cdot
(x-a)^k,\;\;\text{for all}\;\: n\geq 1, \label{150424:0907}
\end{equation}
which then, secondly, will undergo the transformation
$x\mapsto\hat{x}=x-a$ to finally give the equivalent matching
relation\footnote[2]{Note that the coordinate transformation
$x\mapsto\hat{x}=x-a$ is a permissible transformation within the
determination process for the expansion coefficients
$\lambda_{n,k}$ according to \eqref{150423:1617}, simply because
the process itself is not affected by this transformation.}
\begin{equation}
\frac{1}{5^{2^{-n}}}\left[\,\prod_{k=0}^{n-1}
\left(1+\frac{1}{2^{k-2}}\right)^{2^{k-n}}\,\right]\cdot
(\hat{x}+a)^{\frac{1}{2^{n-2}}}=\sum_{k=0}^\infty
\lambda_{n,k}\cdot \hat{x}^k,\;\;\text{for all}\;\: n\geq 1,
\label{150424:1046}
\end{equation}
which, in contrast to \eqref{150424:0907}, is easier to match. In
order to explicitly determine the coefficients $\lambda_{n,k}$
such that equality \eqref{150424:1046} is satisfied for all orders
$n$, it is necessary to expand the power term on the left-hand
side
\begin{equation}
(\hat{x}+a)^\beta=\sum_{k=0}^\infty
\frac{a^{\beta-k}}{k!}\cdot\frac{\Gamma(\beta+1)}{\Gamma(\beta-k+1)}\cdot
\hat{x}^k,\;\;\text{with}\;\; \beta=\frac{1}{2^{n-2}}\geq
0,\;\text{for all}\; n\geq 1. \label{150424:1824}
\end{equation}
Now, since this (transformed) expansion \eqref{150424:1824} only
converges for $|\hat{x}|<|a|$, the original\linebreak
(non-transformed) matching relation \eqref{150424:1819} will
therefore only be valid for $|x-a|<|a|$, with the corresponding
matched coefficients
\begin{equation}
\lambda_{n,k}=\frac{1}{5^{2^{-n}}}\left[\,\prod_{i=0}^{n-1}
\left(1+\frac{1}{2^{i-2}}\right)^{2^{i-n}}\,\right]\cdot
\frac{a^{\frac{1}{2^{n-2}}-k}}{k!}
\cdot\frac{\Gamma\big(\frac{1}{2^{n-2}}+1\big)}{\Gamma\big(\frac{1}{2^{n-2}}-k+1\big)},
\;\; n\geq 1, \;\; k\geq 0.\qquad\square \label{150424:1908}
\end{equation}
Note that since the general solution $y_n$ \eqref{150423:1555} was
matched to a genuine solution of \eqref{150420:0806}, namely to
the special solution $y_n^{(3)}$ \eqref{150423:2018}, and not to
some arbitrary function, the matched coefficients
\eqref{150424:1908} will thus automatically satisfy the
corresponding solved relations \eqref{150423:1617} for $a\neq 0$,
i.e. at least one set of constants $c_n$ for all $n\geq 1$ can be
found which then, according to \eqref{150423:1617}, uniquely
represent the expansion coefficients $\lambda_{n,k}$
\eqref{150424:1908}.

\setcounter{section}{9}

\section{The existence of only two uncoupled Lie point group
invariances\label{J}}

Performing a systematic Lie point group invariance analysis on the
infinite system of first order Riccati-ODEs \eqref{150420:0806},
and looking out only for uncoupled solutions in the overdetermined
system for the generating infinitesimals, which themselves can
then only take the consistent~form
\begin{equation}
\xi(x,y_1,y_2,\dotsc)=\phi(x),\;\;\text{and}\;\;\;
\eta_n(x,y_1,y_2,\dotsc)=\psi_n(x,y_n),\;\;\text{for all}\; n\geq
1,
\end{equation}
one obtains the following infinite recursive set of constraint
equations
\begin{equation}
\left[\frac{\partial\psi_n}{\partial x}x^3+
\left(\frac{\partial\psi_n}{\partial
y_n}-\frac{\psi_n}{y_n}-\frac{d\phi}{dx}+\frac{\phi}{x}\right)y_n
x^2\right]+\left[\left(\frac{\partial \psi_n}{\partial
y_n}-\frac{2\psi_{n+1}}{y_{n+1}}-\frac{d\phi}{dx}+\frac{3\phi}{x}
\right)y_{n+1}^2\right]=0.
\end{equation}
This equation can only be fulfilled if the terms in each of the
two square brackets vanish separately, because, due to that the
first square bracket only depends on $y_n$ and the second one on
$y_{n+1}$, both square brackets are independent of each other. The
above equation thus breaks apart into the following two equations
\begin{align}
\frac{\partial\psi_n}{\partial x}x+
\left(\frac{\partial\psi_n}{\partial
y_n}-\frac{\psi_n}{y_n}-\frac{d\phi}{dx}+\frac{\phi}{x}\right)y_n=0,\label{150519:1335}\\[0.75em]
\frac{\partial \psi_n}{\partial
y_n}-\frac{2\psi_{n+1}}{y_{n+1}}-\frac{d\phi}{dx}+\frac{3\phi}{x}=0.\label{150425:2014}
\end{align}

\noindent The last equation \eqref{150425:2014}, however, is only
consistent if $\psi_n$ is restricted to be a non-shifted linear
function of $y_n$, i.e. if
\begin{equation}
\psi_n(x,y_n)=\alpha_n(x)\cdot y_n,
\end{equation}
which then reduces the system
\eqref{150519:1335}-\eqref{150425:2014} respectively to
\begin{align}
\frac{d\alpha_n}{dx}x-\frac{d\phi}{dx}+\frac{\phi}{x}=0,\label{150519:1927}\\[0.75em]
\alpha_n-2\alpha_{n+1}-\frac{d\phi}{dx}+\frac{3\phi}{x}=0.\label{150519:1930}
\end{align}

\noindent Since \eqref{150519:1927} leads to the result that
$\alpha_n=\alpha_{n+1}$, equation \eqref{150519:1930} gives the
solution for $\alpha_n$ in terms of $\phi$
\begin{equation}
\alpha_n = -\frac{d\phi}{dx}+\frac{3\phi}{x}.\label{150519:2000}
\end{equation}
Inserting this result back into \eqref{150519:1927} leads to the
following differential equation for $\phi$
\begin{equation}
x\frac{d^2\phi}{dx^2}-2\frac{d\phi}{dx}+2\frac{\phi}{x}=0,
\end{equation}
which has the general solution
\begin{equation}
\phi(x)=c_1\cdot x + c_2\cdot x^2,
\end{equation}
which finally, according to \eqref{150519:2000}, implies that
\begin{equation}
\alpha_n(x)=2c_1+c_2\cdot x.
\end{equation}
Hence, the only possible combination in the infinitesimals which
lead to uncoupled Lie point group invariances in the infinite
system \eqref{150420:0806} is given by the 2-dimensional Lie
sub-algebra
\begin{equation}
\xi(x,y_1,y_2,\dotsc)=c_1\cdot x + c_2\cdot x^2,\quad
\eta_n(x,y_1,y_2,\dotsc) = \left(2c_1+c_2\cdot x\right) y_n,\;\;
\text{for all $n\geq 1$}, \label{150522:0049}
\end{equation}
with $[X_1^\infty,X^\infty_2]=X^\infty_2$, where the to
\eqref{150522:0049} corresponding scalar operators $X^\infty_1$
and $X^\infty_2$ are given by \eqref{150424:2258}.

\vspace{3em}\pagebreak[4]
\bibliographystyle{jfm}
\bibliography{BibDaten}

\end{document}